\documentclass{article}

\usepackage{arxiv}

\usepackage[utf8]{inputenc} 
\usepackage[T1]{fontenc}    
\usepackage{url}            
\usepackage{booktabs}       
\usepackage{amsfonts}       
\usepackage{nicefrac}       
\usepackage{microtype}      
\usepackage{lipsum}		
\usepackage{graphicx}
\usepackage{doi}
\usepackage{amssymb}
\usepackage{amsthm}
\usepackage[cmex10]{amsmath}
\usepackage{adjustbox} 
\usepackage{arabtex} 
\usepackage{academicons}
\usepackage{breakcites} 
\usepackage{color}
\usepackage{epsfig}
\usepackage{epstopdf}
\usepackage{etoolbox}
\usepackage{enumitem}
\usepackage{float}
\usepackage{fancyhdr}
\usepackage{graphicx}
\usepackage{graphics}
\usepackage{latexsym,amsfonts}
\usepackage{longtable}
\usepackage{listings}
\usepackage{lscape} 
\usepackage{lipsum}
\usepackage{multirow}             
\usepackage{pdfpages}
\usepackage[figuresright]{rotating} 
\usepackage{sectsty}
\usepackage{setspace}
\usepackage{subfig}
\usepackage{caption}
\usepackage{subcaption}
\usepackage{xcolor,graphicx,subfig,lipsum}
\usepackage{utf8} 
\usepackage{url} 
\usepackage{wrapfig} 
\usepackage{wasysym} 
\usepackage{xcolor}
\usepackage{tabularx}
\usepackage{tikz}
\usetikzlibrary{calc}
\usepackage{tikz-cd}
\usepackage{pifont}
\usetikzlibrary{matrix}
\usepackage{tikz-3dplot}
\usetikzlibrary{patterns}
\usepackage{lmodern}
\usepackage[T1]{fontenc}
\usepackage[sorting=none,bibstyle=ieee,citestyle=numeric-comp,backend=bibtex]{biblatex}
\usepackage{csquotes}
\usepackage{hyperref}
\usepackage{tabu}
\usepackage{adjustbox}
\usepackage{multirow}
\usepackage{multicol}
\usepackage{amsmath,bm}
\usepackage{scalerel,stackengine}
\usepackage[nomain]{glossaries-extra}

\stackMath
\newcommand\reallywidehat[1]{%
	\savestack{\tmpbox}{\stretchto{%
			\scaleto{%
				\scalerel*[\widthof{\ensuremath{#1}}]{\kern-.6pt\bigwedge\kern-.6pt}%
				{\rule[-\textheight/2]{1ex}{\textheight}}
			}{\textheight}%
		}{0.5ex}}%
	\stackon[1pt]{#1}{\tmpbox}%
}
\newcommand{\uperpv}{ \bm{u^{\perp} } }
\newcommand{\uperp}{ u^{\perp}  }
\newcommand{\nabperp}{\nabla^{\perp}}
\newcommand{\deriv}[3][]{
	\ensuremath{\frac{\partial^{#1} {#2}}{\partial {#3}^{#1}}}}

\title{A Hybrid Discrete Exterior Calculus Discretization and Fourier
Transform of the Incompressible Navier-Stokes Equations in 3D}

\author{ \href{https://orcid.org/0000-0002-5216-6753}{\includegraphics[scale=0.06]{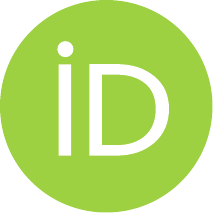}\hspace{1mm}Abdullah Abukhwejah}\thanks{Electronic mail: abdullah.abukhwejah@kaust.edu.sa}, \href{https://orcid.org/0000-0002-5033-1970}{\includegraphics[scale=0.06]{orcid.pdf}\hspace{1mm}Pankaj Jagad}\thanks{Electronic mail: pankajjagad9670@iitbombay.org}, \href{https://orcid.org/0000-0002-4702-6473}{\includegraphics[scale=0.06]{orcid.pdf}\hspace{1mm}Ravi Samtaney}\thanks{This paper is dedicated to the memory of Prof. Ravi Samtaney, who passed away while this work was in progress.} and \href{https://orcid.org/0000-0002-2257-8490}{\includegraphics[scale=0.06]{orcid.pdf}\hspace{1mm}Peter Schmid} \\
	Mechanical Engineering, Physical Science and Engineering Division\\
	King Abdullah University of Science and Technology, Thuwal,\\
	Saudi Arabia\\
}

\renewcommand{\shorttitle}{Hybrid DEC-FT Discretization of NSE}

\hypersetup{
pdftitle={A Hybrid Discrete Exterior Calculus Discretization and Fourier
Transform of the Incompressible Navier-Stokes Equations in 3D},
pdfsubject={q-bio.NC, q-bio.QM},
pdfauthor={Abdullah Abukhwejah, Pankaj Jagad, Ravi Samtaney, Peter Schmid},
pdfkeywords={Discrete Exterior Calculus, Navier-Stokes Equations, Fourier
Transform, Discretization},
}

\begin{document}
\maketitle

The simulation of fluid flow problems, specifically incompressible flows governed by the Navier-Stokes equations (NSE), holds fundamental significance in a range of scientific and engineering applications. Traditional numerical methods employed for solving these equations on $three{\text -}dimensional$ (3D) meshes are commonly known for their moderate conservation properties, high computational intensity and substantial resource demands. Relying on its ability to capture the intrinsic geometric and topological properties of simplicial meshes, discrete exterior calculus (DEC) provides a discrete analog to differential forms and enables the discretization of partial differential equations (PDEs) on meshes. We present a hybrid discretization approach for the 3D incompressible Navier-Stokes equations based on DEC and Fourier transform (FT). An existing conservative primitive variable DEC discretization of incompressible Navier–Stokes equations over surface simplicial meshes developed by Jagad et al. \cite{jagad2021primitive} is considered in the planar dimension while the Fourier expansion is applied in the third dimension. The test cases of three-dimensional lid-driven cavity and viscous Taylor-Green three-dimensional vortex (TGV) flows show that the simulation results using this hybrid approach are comparable to literature.

\keywords{Discrete Exterior Calculus \and Navier-Stokes Equations \and Fourier
Transform \and Discretization}



\section{Background} 
Simulating fluid flow problems, particularly incompressible flows described by the Navier-Stokes equations (NSE), is essential for numerous scientific and engineering applications. Conventional numerical methods for solving these equations on $three{\text -}dimensional$ (3D) meshes are often computationally intensive and resource-demanding. Discrete exterior calculus (DEC) is a powerful mathematical framework that has been employed to discretize and solve partial differential equations, including the incompressible Navier-Stokes equations. DEC provides a systematic approach to discretizing these equations on a discrete mesh, preserving important mathematical properties. By leveraging concepts from differential geometry, DEC represents the differential operators in the Navier-Stokes equations as discrete operators on simplicial complexes. This allows for the formulation of discrete counterparts of the divergence, gradient, and curl operators, which are key components of the Navier-Stokes equations. By discretizing the equations using DEC, one can simulate and analyze fluid flow phenomena with improved accuracy and computational efficiency, unveiling new possibilities for studying complex fluid dynamics problems.\\
DEC is a rapidly growing field with many potential applications. In computational fluid dynamics, DEC can be used to simulate the flow of fluids such as darcy flow \cite{hirani2015numerical}, lid-driven cavity flow \cite{mohamed2016discrete}, flow past as circular cylinder \cite{jagad2018flow}, and flow over an airfoil \cite{jagad2021primitive}. DEC can also be used to create realistic models of surfaces and objects in the field of computer graphics and computer vision \cite{mollenhoff2019lifting, desbrun2003discrete, grinspun2006discrete, elcott2007stable, mullen2009energy, crane2013digital, de2016subdivision}.
Moreover, DEC is used in medical imaging \cite{wei2010interactive} to reconstruct the shape of organs and tissues from medical scans and to compute the complex blood flow patterns within the aneurysm.
In robotics, DEC can be used to design and control robots \cite{kupcsik2021supervised}. In addition to that,  the application of DEC is extended to materials science to study the properties of materials \cite{desbrun2005discrete}. Electromagnetic theory is developed with DEC \cite{chen2017electromagnetic} and led to satisfaction of Gauss$\text{'}$  Theorem and Stokes$\text{'}$ Theorem. Thus, DEC is a versatile tool with many potential applications.
\section{Motivation} 
\subsection{Discrete Exterior Calculus}
\begin{figure}
		\centering
		\includegraphics[width=0.55\columnwidth]{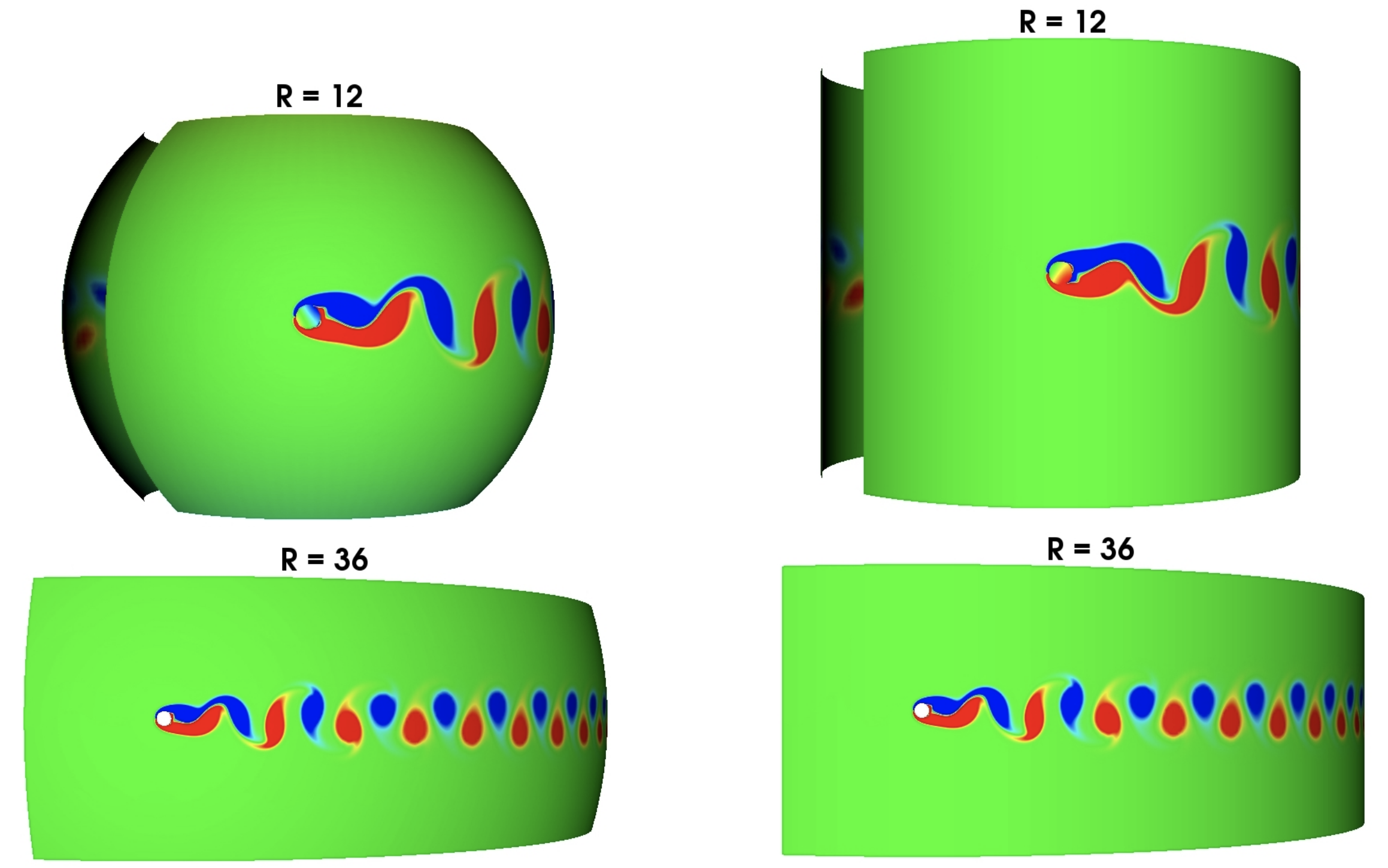}
		\caption{Vortex streets for flow past a circular cylinder embedded on a spherical surface (left) and a cylinderical surface (right). $R$ represents the radius of the embedding surface. \cite{jagad2018flow}.}
		\label{deccoord}
	\end{figure}

Utilizing structure-preserving numerical methods is of paramount importance in computational science and engineering as it ensures that the mathematical and physical integrity of governing equations is precisely retained throughout the discretization process. The significance lies in the preservation of critical properties, leading to more accurate and reliable simulations of real-world phenomena. DEC emerges as a standout method in this context. By utilizing integral quantities and discrete forms as primary unknowns, DEC converts continuous partial differential equation (PDE) systems into discrete algebraic equivalents. What sets DEC apart is its unique capability to confine errors and approximations solely to the algebraic level, avoiding deterioration in the approximation of the PDEs' differential operators. This distinctive characteristic makes DEC a powerful tool in maintaining the underlying physical and mathematical properties of continuous systems, ensuring that numerical solutions remain accurate representations of the modeled phenomena. In essence, DEC stands as a testament to the importance of structure-preserving methods, providing a robust framework for accurate computational modeling in diverse scientific and engineering applications.\\
\\
Exterior Calculus (EC) is a relatively recent approach for solving partial differential equations, rooted in the concept of discretizing the mathematical framework of Exterior Differential Calculus as formalized by Cartan \cite{cartan1899certaines}. In his PhD thesis, Hirani \cite{hirani2003discrete} formulated the foundation of DEC using discrete combinatorial and geometric operations on simplicial complexes. In DEC, the discrete operators are computed to be used in numerical methods for solving partial differential equations PDEs. The method is mimetic and thus many of the rules/identities of its continuous counterparts are retained at the discrete level \cite{hirani2003discrete,desbrun2005discrete}. This leads to enhanced conservation properties for DEC discretization of physical problems. The DEC numerical method adheres to orthogonality relations through the incorporation of differential forms; $\nabla \times {\bf v}=0$ for any vector $\bf v$ and $\nabla \times {\nabla \bf s}=0$ for any scalar $\bf s$. A key property of DEC is its ability to operate in arbitrarily high dimensions and its independence from any specific coordinate system \cite{perot2014differential,jagad2018flow}, i.e., applying DEC for solving physical problems on diverse embedding surfaces can be achieved using the identical set of equations, as demonstrated in Figure \ref{deccoord}, without any need for adjustments or modifications. 
 \\
 The application of DEC for solving PDE was first introduced by Hirani et al. \cite{hirani2015numerical} to solve Darcy flow and Poisson's equation. 
 By leveraging on the DEC unique features of coordinate independent and structures preservation, Mantravadi et al. \cite{mantravadi2023hybrid} recently developed a hybrid discrete exterior calculus and finite difference (DEC-FD) method for Boussinesq thermal convection in spherical shells and then the method was verified and extended for anelastic convection in spherical shells by Jagad et al. \cite{jagad2023anelastic} and Khan et al. \cite{khan2023hybrid}. 
 
 Navier-Stokes equations were first rephrased in the stream function formulation and solved using DEC by Mohamed et al. \cite{mohamed2016discrete} for incompressible flows over $two{\text -}dimensional$ (2D) flat/curved domains. Nitschke et al. \cite{nitschke2017discrete} presented a discretization of the surface Navier-Stokes equation using DEC which is capable of handling Gaussian curvature and dealing with harmonic vector fields. Then, based on adopting the primitive variable formulation and incorporating the Gaussian curvature term to consider the curvature of the embedding surface, Jagad et al. \cite{jagad2018flow,jagad2018vortex} investigated the vortex dynamic and flow past a stationary circular cylinder embedded on spherical and cylindrical 2D surfaces using DEC at a fixed Reynolds number of ($Re$) $100$. A primitive variable DEC discretization of incompressible NSEs incorporating energy-preserving time integration and the Coriolis force was presented by \cite{jagad2021primitive} aiming to broaden its applicability for studying the late-time behavior of flows on rotating surfaces. Wang et al. \cite{wang2023discrete} presented a DEC discretization of two-phase incompressible Navier-Stokes equations with a conservative phase field method and various applications such as Rayleigh-Taylor instability, dam breaking, and rising bubble on flat and curved domains were considered.

 \subsection{Fourier Transform}
The Fourier transform is a powerful mathematical tool that has been widely used in solving the Navier-Stokes equations. The Fourier transform allows for the decomposition of a function into its frequency components, which can be useful in analyzing and solving differential equations.  One application of the Fourier transform in solving the Navier-Stokes equations is in the simulation of wall-bounded two-phase flows. Challa et al. \cite{challa2018three} presented a hybrid spectral element-Fourier spectral method for solving the coupled system of Navier-Stokes and Cahn-Hilliard equations in a 3D domain. The Fourier spectral method is used to handle the homogeneous direction by applying  fast Fourier transforms (FFT), while the spectral element method is used for the other directions. This hybrid approach allows for an efficient and accurate simulation of wall-bounded two-phase flows. The Fourier transform can also be used in the analysis of the Navier-Stokes equations for shear flows. Moser et al. \cite{moser1983spectral} developed a spectral method for the Navier-Stokes equations with applications to Taylor-Couette flow based on Chebychev polynomials and Fourier transforms. Furthermore, the Fourier transform has been used in the numerical simulation of turbulent flows. Ma et al. \cite{ma2000dynamics} investigated the dynamics and low-dimensionality of a turbulent near wake using large eddy simulation (LES) and direct numerical simulation (DNS). The Fourier transform is used to analyze the turbulent structures and their interactions, providing valuable information for understanding turbulent flows. In summary, the Fourier transform is a valuable tool in solving the Navier-Stokes equations. It allows for the decomposition of functions into frequency components, which can be used to analyze and solve differential equations. Starn \cite{starn2001simple} introduced an unconditionally stable simple fluid solver based on FFT. A splitting method was introduced by Br{\"u}ger et al. \cite{bruger2005splitting} in which, the solution is expanded in the third dimension by FFT  and the efficiency of the method were demonstrated in a numerical experiment with rotated Poiseuille flow perturbed by Orr-Sommerfield modes in a channel.

\section{Summary of Present Work \label{ResearchSummary}} 
This paper proposes to investigate the extension of DEC and its combination with Fourier transform to provide an efficient approach for solving the incompressible Navier-Stokes equations in 3D, which can significantly reduce computational complexity while maintaining accuracy.\\
The primary objectives of this research are as follows:
\begin{itemize}
	\item Extend the application of DEC to solve the incompressible Navier-Stokes equations.
        \item Integrate the Fourier Transform method into the DEC framework to efficiently handle spatial derivatives in the third dimension.
        \item Assess the accuracy and computational efficiency of the proposed approach and compare it to literature findings.
\end{itemize}
These objectives are achieved by developing a DEC-based formulation for incompressible Navier-Stokes equations in 2D. Then, integrating the Fourier Transform method into the DEC-based formulation to efficiently handle spatial derivatives of the Navier-Stokes equations in the third dimension. Eventually, numerical experiments involving complex 2D fluid flow problems, comparing the hybrid DEC-FT approach against traditional 3D mesh-based methods to assess the computational efficiency and accuracy of the proposed method.\\
The proposed research has the potential to significantly reduce computational demands in simulating incompressible flows, allowing for faster and more accessible solutions.
It contributes to advancing the field of DEC by extending its application to fluid dynamics in a novel way.
This research has practical applications in engineering, environmental sciences, and computational fluid dynamics, enabling more efficient and accurate simulations of fluid flow problems.

\section{Paper Outline} 
The paper is outlined as follows. An overview of the present research work is presented in Section \ref{ResearchSummary}. Sections \ref{GoverningEquations} and \ref{ECnotation} include the governing equations and their expression in exterior calculus notation, respectively. Section \ref{DiscretizationMethod} gives the details of the discretization and the present hybrid DEC-FFT method. In Sections \ref{LDCProblemFormulation} and \ref{LDCValidation}, implementation of the hybrid method within the context of 3D lid-driven cavity flow is explored and comparisons with previously published results in the literature are drawn. Sections \ref{2DTGV} and \ref{3DTGV} delve into the utilization of the hybrid method to analyze 2D and 3D Taylor-Green vortices flows, offering a comparative analysis against existing literature findings. The key results derived from this work and potential areas for future exploration are encapsulated in the concluding Summary Section  \ref{Summary}.





\section{Governing Equations \label{GoverningEquations}} 
For fluid flow with unit density and in the absence of body forces, the incompressible Navier-Stokes equations are expressed as follows:
\begin{equation}
\frac{\partial{\bf u}} {\partial t} - \mu \nabla^2 {\bf u} + {(\bf u \cdot \nabla) u} +  {\bf \nabla} p = 0
\label{eq1}
\end{equation}
\begin{equation}
\nabla \cdot {\bf u}=0
\label{incomp}
\end{equation}
Here, ${\bf u}$ represents the velocity vector, $p$ denotes the pressure, and $\mu$ stands for the dynamic viscosity.\\

\section{Navier-Stokes Equations in Exterior Calculus Notation \label{ECnotation}}

In the initial stage of discretization, it is essential to transform the Navier-Stokes equations from vector calculus notation to exterior calculus notation. This conversion is pivotal as it aligns the equations with the principles of DEC, facilitating a more systematic approach to the numerical method. Initially, the process involves substituting the equations with tensor identities:\\
\begin{equation}
\nabla^2 {\bf u} = \nabla (\nabla \cdot {\bf u}) - \nabla \times (\nabla \times {\bf u} )
\end{equation}
\begin{equation}
{(\bf u \cdot \nabla) u} = \frac{1}{2} \nabla ({\bf u} \cdot {\bf u}) - {\bf u} \times (\nabla \times {\bf u})
\end{equation}
\\
and incorporating the constraint of incompressibility $\nabla \cdot {\bf u}=0$, equation \ref{eq1} can be expressed as
\begin{equation}
\frac{\partial{\bf u}} {\partial t} + \mu \space  \nabla \times \nabla \times {\bf u} - {\bf u} \times (\nabla \times {\bf u} ) +  \nabla p^d = 0
\label{ns-rota}
\end{equation}
where $p^d$  denotes the dynamic pressure defined as $p^d = p + \frac{1}{2} ({\bf u} \cdot {\bf u} ) $
\\
The notation transformation involves applying the flat operator $(\flat)$ to Equations \ref{eq1} and \ref{incomp}, and then substituting with the corresponding identities:
\begin{equation}
(\nabla \times \nabla \times {\bf u})^\flat = (-1)^{(N+1)} \ast d \ast d{\bf u}^\flat
\end{equation}
\begin{equation}
({\bf u} \times (\nabla \times {\bf u}))^\flat = (-1)^{(N+1)} \ast ( {\bf u}^\flat \wedge \ast d{\bf u}^\flat  )
\end{equation}
\begin{equation}
(\nabla \cdot {\bf u})^\flat = \ast d \ast {\bf u}^\flat
\end{equation}
\begin{equation}
(\nabla p^d)^\flat = dp^d
\end{equation}

\begin{flushleft} 
	where
	\begin{itemize}
		\item $\ast$ denotes the hodge star operator
		\item $d$ denotes the exterior derivative operator
		\item $\wedge$ denotes the wedge product operator
		\item $\flat$ denotes the flat operator which transforms a vector $\bf u$ into a 1-form $\bf u^\flat$
		\item $N$ denotes the space dimension
	\end{itemize}
\end{flushleft}

Substituting previous identities in Eq. \ref{ns-rota} and \ref{incomp}, N-S equations can be expressed as

\begin{equation}
\frac{\partial{\bf u}^\flat} {\partial t} + (-1)^{(N+1)} \mu \ast d \ast d{\bf u}^\flat + (-1)^{(N+2)} \ast ( {\bf u}^\flat \wedge \ast d{\bf u}^\flat  ) + dp^d = 0
\label{ns-d1}
\end{equation}
\begin{equation}
\ast d \ast {\bf u}^\flat=0
\label{ns-d2}
\end{equation}
\\
where the velocity field is expressed as the 1-form ${\bf u}^\flat$, and $p^d$ denotes the dynamic pressure as a 0-form. These equations represent the vorticity form of Navier–Stokes equations denoted in exterior calculus notation. Full details can be found in \cite{mohamed2016discrete}.

Considering a $two{\text -}dimensional$ space and omitting the flat operator for convenience, Equations \ref{ns-d1} and \ref{ns-d2} take the form

\begin{equation}
\frac{\partial {\bf u}}{\partial t}  - \mu \ast d \ast d {\bf u}  + \ast ({\bf u} \wedge \ast d {\bf u} ) + dp^d= 0
\label{nm-0}
\end{equation}
\begin{equation}
\ast d \ast  {\bf u}  = 0
\label{nm-1}
\end{equation}

which represent the Navier–Stokes equations formulated in exterior calculus notation using primitive variables.

\section{The Discretization Method \label{DiscretizationMethod}}
This section starts by introducing the notation for the simplicial mesh used to discretize the simulation domain. Following this, a brief overview of discrete differential forms and some discrete operators is presented. The derivation of the discretization of NS equations is then presented for 2D. In this context, the discussion on the simplicial mesh and discrete operators is intentionally concise. Further elaboration and details can be found in \cite{hirani2003discrete,desbrun2005discrete,hirani2015numerical}.
\subsection{The Domain Discretization}
Let $\Omega$ be the physical domain with a dimension $N = 2$, approximated by the simplicial complex $K$ as described in \cite{hirani2003discrete,hirani2015numerical}. A simplex $\sigma$ within the domain has dimension $k$, represented as $\sigma^k \in K$. The nodes define a $k$-simplex $\sigma^k = [v_0, \dots, v_k]$, where the subscripts denote the indices of the nodes. The sequence in which nodes are arranged to form a simplex determines its orientation. It is assumed that the top-dimensional simplices $\sigma^N$ have been consistently oriented, whereas the orientation of lower-dimensional simplices is arbitrary.
 
\begin{figure}
		\centering
		\includegraphics[width=0.55\columnwidth]{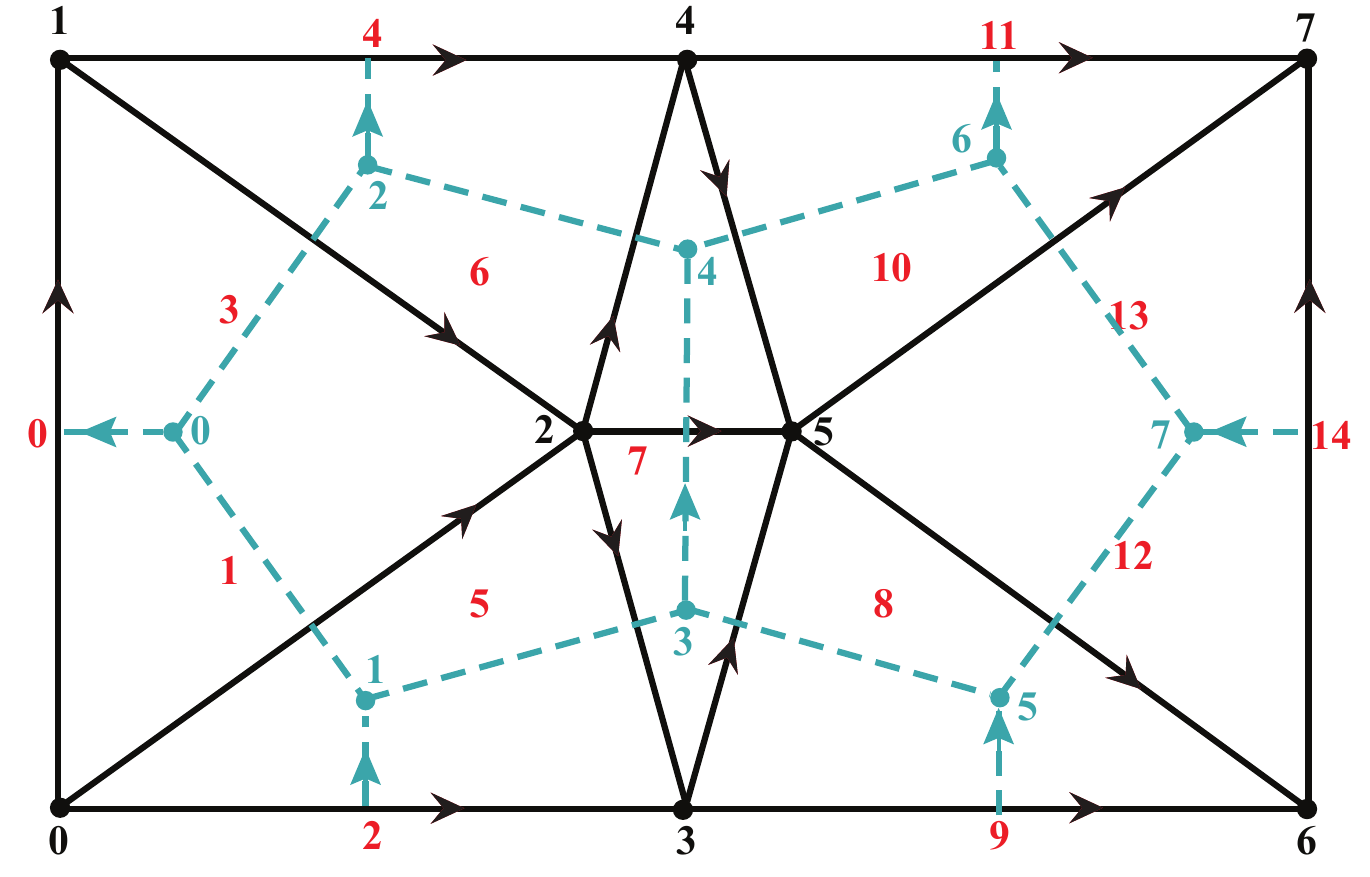}
		\caption{A representation of a 2D simplicial mesh depicting the primal simplices (in black) and their corresponding dual cells (in green).}
		\label{simpl}
	\end{figure}

Figure \ref{simpl} shows an example of 2D simplicial mesh. In this mesh, the number of $k$-simplices is denoted by $N_k$, i.e., $N_2 = 8$, $N_1 = 15$ and $N_0 = 8$. Dual complex $\star K$ is associated with the primal simplicial complex $K$. The dual of a primal $k$-simplex $\sigma^k \in K$ is represented as an $(N-k)$-cell denoted by $\star \sigma^k \in K$. The circumcentric dual is considered for the dual mesh shown in Figure \ref{simpl} in green color. In the context of a 2D mesh, the circumcenter serves as the dual of a triangle, while the dual of a primal edge is a connecting dual edge between the circumcenters of adjacent triangles. Similarly, the dual of a primal node manifests as a 2-cell polygon comprising the duals of the connected edges. In scenarios where a triangular mesh represents a curved surface, the dual edges can exhibit kinks, and the dual cells may not lie on a single plane. In this 2D setting, both primal triangles and dual polygons are assumed to possess a counterclockwise orientation. Although the orientation of primal edges is arbitrary, their orientations dictate those of the dual edges, which can be established by rotating the orientation of the primal edge 90 degrees counterclockwise, as depicted in Figure \ref{simpl}.

\subsection{Discrete Exterior Calculus}
Discrete exterior calculus offers discrete formulations for numerous exterior calculus operators, including the exterior derivative, Hodge star, wedge product, and other related operators \cite{hirani2003discrete,desbrun2005discrete}.These distinct operators possess the benefit of adhering to numerous rules and identities akin to their continuous counterparts. The mimetic properties of these discrete operators are known for maintaining the underlying physics as implied by the continuous governing equations when transitioning to discrete form \cite{perot2007discrete}, which subsequently enhances the accuracy of the numerical discretization method in representing the physical fidelity.

The space of discrete $k$-forms on the primal mesh complex is denoted by $C^k(K)$, while the space of discrete $k$-forms on the dual mesh complex is denoted by $D^k(\star K)$. The discrete exterior derivative and Hodge star operators establish connections between these spaces, as illustrated in the diagram (\ref{cochain}).

	\begin{equation}
	\begin{tikzcd}
	C^0(K) \arrow{r} {d_0} \arrow[d, xshift=0.7ex, "{\ast_0}"]
	& C^1(K) \arrow[d, xshift=0.7ex, "{\ast_1}"] \arrow{r} {d_1}  &  C^2(K) \arrow[d, xshift=0.7ex, "{\ast_2}"]  \\
	D^2(\star K)  \arrow [u, xshift=-0.7ex, "{\ast^{-1}_0}"]] 
	& D^1( \star K) \arrow{l} {{-d_0}^T} \arrow[u, xshift=-0.7ex, "{\ast^{-1}_1}"]  &    D^0(\star K) \arrow{l} {{d_1}^T} \arrow[u, xshift=-0.7ex, "{\ast^{-1}_2}"] 
	\end{tikzcd}
	\label{cochain}
	\end{equation}
 The discrete exterior derivative operator $d_k$ transforms primal 
$k{\text -}$forms into primal 
$(k+1){\text -}$forms, while the counterpart that converts dual $k{\text -}$forms into dual $(k+1){\text -}$forms is derived from the transpose of the $d(N-k-1)$ operator. In 2D, this adaptation involves introducing a negative sign solely for the
$d^{T}_0$ operator due to the predefined mesh orientation convention. Additionally, the discrete Hodge star operator $\ast_k$ maps primal $k$-forms to dual $(N-k)$-forms, along with its inverse counterpart, denoted as $\ast^{-1}_k$, performing the reverse transformation from dual $(N-k){\text -}$forms back to primal $
k{\text -}$forms.

The discrete exterior derivative operator $d_k$ is represented by a sparse matrix, which is essentially the transpose of the boundary operator applied to the $(k+1){\text -}$simplices. For instance, in the context of a 2D mesh as illustrated in Figure \ref{simpl}, the discrete $d_1$ operator, responsible for mapping primal 1-forms defined on the edges of the primal mesh to 2-forms defined on the triangles of the mesh, takes the form of an $N_2 \times N_1$ matrix, characterized as follows:
\begin{equation}
    {[d_1]}_{ij} = 
    \begin{cases}
      +1 & \text{if edge $j$ constitutes one of the boundaries of triangle $i$,} \\
       & \text{and they maintain consistent orientations}\\
      -1 & \text{if edge $j$ constitutes one of the boundaries of triangle $i$,} \\
       & \text{and they do not maintain consistent orientations}\\
      0 & \text{if edge $j$ does not constitute one of the boundaries of triangle $i$}
    \end{cases}     
    \label{discretederivative}
\end{equation}
\\

For primal nodes located on the domain boundary, such as node 4 in Figure \ref{simpl}, the boundary of their corresponding dual 2-cells (polygons) encompasses primal boundary edges. Consequently, the $[-d^T_0]$ matrix, depicted by the transposed boundary operator of these dual 2-cells, is augmented by another operator that accommodates the primal boundary edges.

The discrete Hodge star operator $\ast_k$ is essentially a diagonal matrix. Each element on the diagonal corresponds to the ratio of the volume of the dual $(N - k){\text -}$cell $\star \sigma^k_i$ to the volume of its primal k-simplex $\sigma^k_i$, denoted as $\frac{|\star\sigma^k_i|}{|\sigma^k_i|}$. As for the wedge product operator, its discrete definition will be outlined in the following two subsections within the context of discretization demonstration.

\subsection{Two Dimensional Discretization of Navier-Stokes Equations}

For the domain discretization using the primal mesh and its corresponding dual mesh, Equation (\ref{nm-0}) in DEC notation is expressed as
\begin{equation}
\frac{{U^{n+1}} - {U^{n}}}{\Delta t} - \mu \ast_1 d_0  \ast^{-1}_0 \big[  [- d^{T}_0] U + d_b V \big] + \ast_1 W_v \ast^{-1}_0 \big[  [- d^{T}_0] U + d_b V \big] + d^{T}_1 P^d = 0
\label{nm-2}
\end{equation}

where $U$, $V$, and $P^d$ are vectors representing the discrete dual normal velocity 1-forms for all dual edges of the mesh, the primal tangential velocity 1-forms for all primal edges of the mesh, and the dynamic pressure dual 0-forms for all dual vertices of the mesh, respectively. The matrix $W_v$ represents the discrete wedge product of the tangential velocity 1-form $v$ with the 0-form $\ast du$, and it contains the values of the tangential velocity 1-form $v$. 
The operation $[-d^{T}_0]U$ computes the circulation of the velocity forms $u$ along the boundaries of the dual 2-cells. The operation $d_bV$ complements this circulation by considering the parts of the dual 2-cell boundaries that consist of primal edges, thereby accounting for the contribution from the velocity 1-forms $v$ on these primal boundary edges. Define $U^{\ast}$ as the vector comprising the mass flux primal 1-form $u^{\ast}$ for all primal edges of the mesh, which can be expressed as $U^{\ast} = -\ast^{-1}_1 U$ and $U = \ast_1 U^{\ast}$. Equation (\ref{nm-2}) becomes

\begin{equation}
\ast_1 \frac{{U^{n+1}} - {U^{n}}}{\Delta t} - \mu \ast_1 d_0  \ast^{-1}_0 \big[  [- d^{T}_0] \ast_1 U^{\ast} + d_b V \big] + \ast_1 W_v \ast^{-1}_0 \big[  [- d^{T}_0] \ast_1 U^{\ast} + d_b V \big] + d^{T}_1 P^d = 0
\label{nm-3}
\end{equation}

By applying the Hodge star operator $\ast^{-1}_1$ to Equation (\ref{nm-3}) and utilizing the property $\ast^{-1}_1 \ast_1 = -1$, Equation (\ref{nm-3}) transforms into

\begin{equation}
- \frac{{U^{n+1}} - {U^{n}}}{\Delta t} + \mu d_0  \ast^{-1}_0 \big[  [- d^{T}_0] \ast_1 U^{\ast} + d_b V \big] - W_v \ast^{-1}_0 \big[  [- d^{T}_0] \ast_1 U^{\ast} + d_b V \big] + \ast^{-1}_1 d^{T}_1 P^d = 0
\label{nm-4}
\end{equation}

The continuity equation can be written as 
\begin{equation}
[d_1] U^{\ast} + [0]P^d = 0
\label{nm-5}
\end{equation}

More details on incorporating Euler's first-order time integration and an energy-preserving second order time integrator can be found in \cite{jagad2021primitive}. The latter can be considered in order to perform simulations at higher Reynolds numbers.
\subsection{Three Dimonsional Hybrid DEC-FFT Discretization of NSEs}
The hybrid DEC-FFT method is based on obtaining the solutions of velocity and pressure in 2D by DEC, and expanding them in the third dimension by FFT. Fourier expansion is obtained by assuming periodic boundary conditions. As shown above for the momentum and continuity equations, FFT is taken for each term in order to obtain a complete set of velocity and pressure solutions in the span-wise direction.

	 For fluid flow with unit density and no body forces, the incompressible Navier-Stokes equations are given by
	\begin{equation}
		\frac{\partial \bm{u}}{\partial t} + \bm{u} \cdot \nabla \bm{u} = - \nabla p + \mu \nabla^2 \bm{u}
		\label{mom0}
	\end{equation}
	
	\begin{equation}
		\nabla \cdot \bm{u} = 0
		\label{cont0}
	\end{equation}
	
	where $\bm{u}$ is the velocity vector, $p$ is the pressure, and $\mu$ is the dynamic viscosity.\\

		 DEC is considered in the perpendicular ($\perp$), $xy$ plane and the Fast Fourier Transform (FFT) is taken only in the $z$-direction.
		 Let $\bm{u} = (\uperpv, w)$ where $\uperpv$ is the velocity vector in the $xy$ plane and $w$ represents the velocity in the $z$-direction. 
		 The gradient operator can be defined as $\nabla = (\nabla^{\perp}, \frac{\partial}{\partial z})$ where $ \nabla^{\perp} = ( \frac{\partial}{\partial x}, \frac{\partial}{\partial y} )$.

		 Substituting the velocity vector and the differential operators, Equation (\ref{cont0}) can be expressed as
		\begin{equation}
			\nabla^{\perp} \cdot \uperpv + \frac{\partial w}{\partial z} = 0
			\label{cont_ec}
		\end{equation}
		
		 Using the flat operator, the velocity 1-form can be obtained as $\uperp = ({\uperpv})^{\flat}$. In the EC notation, Equation (\ref{cont_ec}) reads
		\begin{equation}
			\ast \ d \ast  u^{\perp} 	+  \frac{\partial w}{\partial z}  = 0
		\end{equation}
Note that the relation between the Fourier transform of a function and its derivative is:
\begin{equation}
\frac{\widehat{\partial \bf u}}{\partial z} = -ik \widehat{u}
\end{equation}

   Taking FFT in the $z{\text -}$direction
		\begin{equation}
			\ast \ d \ast  \widehat {u^{\perp}} 	+  i \ k \ \widehat {w}  = 0
			\label{cont_FT}
		\end{equation}
		
		where $k$ denotes the wave number.

    By splitting into $\perp$ and $z$ components, Equation (\ref{mom0}) can be expressed in $\perp$-momentum equation and $z$-momentum equation as
		 \begin{equation}
		 	\frac{\partial \uperpv}{\partial t} + \uperpv \cdot \nabla^{\perp} \uperpv + w \frac{\partial \uperpv}{\partial z} = - \nabla^{\perp} p + \mu({\nabla^{\perp}}^2 \uperpv + \frac{\partial^2 \uperpv}{\partial z^2})
		 	\label{momperp}
		 \end{equation}
		 \begin{equation}
		 	\frac{\partial w}{\partial t} + \uperpv \cdot \nabla^{\perp} w + w \frac{\partial w}{\partial z} = - \frac{\partial p}{\partial z} + \mu ( {\nabla^{\perp}}^2 w + \frac{\partial^2 w}{\partial z^2} )
		 	\label{z_mom_vec}
		 \end{equation}

		 using tensor identities along with continuity equation, nonlinear and viscous terms in Equations (\ref{momperp}) and (\ref{z_mom_vec}) can be rewritten as
	\begin{itemize}
    \item Nonlinear and viscous terms in $\perp$-momentum equation	 
		 \begin{equation}
		 	\uperpv \cdot \nabla^{\perp} \uperpv = \frac{1}{2} \nabla^{\perp} |\uperpv|^2 - \uperpv \times \nabla^{\perp} \times \uperpv
		 \end{equation}
		 \begin{equation}
		 	\begin{aligned}
		 		{\nabla^{\perp}}^2 \uperpv &= \nabla^{\perp} ( \nabla^{\perp} \cdot \uperpv ) - \nabla^{\perp} \times \nabla^{\perp} \times \uperpv \\
		 		&= - \nabla^{\perp} (\frac{\partial w}{\partial z} ) - \nabla^{\perp} \times \nabla^{\perp} \times \uperpv
		 	\end{aligned}
		 \end{equation}
		 
		\item Nonlinear term in $z$-momentum equation
		 \begin{equation}
		 	\begin{aligned}
		 		\uperpv \cdot \nabperp w &= \nabperp \cdot (\uperpv w) - w \nabperp \cdot \uperpv \\
		 		&= \nabperp \cdot (\uperpv w) + w \deriv{w}{z}
		 	\end{aligned}
		 \end{equation}
	\end{itemize}	 
		 
		 substituting into Equation (\ref{momperp}) yields
	\begin{equation}
	\begin{aligned}
	\deriv[]{\uperpv}{t} - \uperpv \times \nabperp \times \uperpv + w \deriv[]{\uperpv}{z} = - \nabperp ( p + \frac{1}{2} |\uperpv|^2 )  \\ - \mu  \Big( \nabperp \times \nabperp \times \uperpv + \deriv[]{}{z} \nabperp w  - \deriv[2]{\uperpv}{z}\Big)
	\end{aligned}
	\label{p_mom_vec}
	\end{equation}
 
	where 	 $p + \frac{1}{2} |\uperpv|^2 = p^d$ and $\uperp$ is a 1-form in the $\perp$ plane.
 The notation transformation in 2D is carried out by applying the following identities:
		 \begin{equation}
		 	\uperpv \times \nabperp \times \uperpv = - \ast (\uperp \wedge \ast d \uperp)
		 \end{equation}
		 \begin{equation}
		 	\nabperp \times \nabperp \times \uperpv = - \ast d \ast d \uperp
		 \end{equation}
   		 \begin{equation}
		 	\nabperp \cdot \uperpv= \ast \ d \ast \uperp
		 \end{equation}
		 \begin{equation}
		 	\nabperp p= d p
		 \end{equation}

		 	$\perp$-momentum equation can be expressed in EC notation as 
		\begin{equation}
		\deriv[]{\uperp}{t} + \ast (  \uperp \wedge \ast d \uperp) + w \wedge \deriv[]{\uperp}{z} = - d p -  \frac{1}{2} d |\uperpv|^2 + \mu \Big(  \ast d \ast d \uperp - \deriv[]{}{z} d w + \deriv[2]{\uperp}{z}   \Big)
		\label{p_mom_ec}
		\end{equation}
		 	
		 similarly, rewriting $z$-momentum equation as
            \begin{equation}
		 		\deriv[]{w}{t} + \nabperp \cdot (\uperpv w) + \deriv[]{}{z} w^2 = - \deriv[]{p}{z} + \mu ( \nabperp \cdot ( \nabperp w) + \deriv[2]{w}{z} )
		 	\label{z_mom_vec2}
		 \end{equation}
		 	
		 	and can be expressed in EC notation as
		 	\begin{equation}
				\deriv[]{w}{t} + \ast d \ast ( \uperp \wedge w ) + \deriv[]{}{z} w^2 = - \deriv[]{p}{z} + \mu \Big(  \ast d \ast d w + \deriv[2]{w}{z}  \Big)
				\label{z_mom_ec}
				\end{equation}
		 	
		 	where $w$ is defined as a 0-form in the $\perp$ plane.

		 	Taking FFT of Equations (\ref{p_mom_ec}) and (\ref{z_mom_ec}) yields
	\begin{equation}
	\deriv[]{\widehat{\uperp}}{t} + \ast \reallywidehat{ (\uperp \wedge \ast d \uperp) } + i k \reallywidehat{ w \wedge \uperp } = - d \widehat{p} -  \frac{1}{2} d \widehat{|\uperpv|^2} + \mu \Big(  \ast d \ast d \widehat{\uperp} - i k d \widehat{w} - k^2 \widehat {u^{\perp}}      \Big)
	\label{mom_perp_FT}
	\end{equation}
	\begin{equation}
	\deriv[]{\widehat{w}}{t} +  \ast d \ast \reallywidehat{(\uperp \wedge w) } + i k \widehat{w^2} = - i k \widehat{p} + \mu \Big( \ast d \ast d \widehat{w} - k^2 \widehat{w}  \Big)
	\label{mom_z_FT}
	\end{equation}
		 	
		 	In addition to the continuity equation which was obtained earlier as:
		 	\begin{equation}
		 		\ast \ d \ast  \widehat {u^{\perp}} 	+  i \ k \ \widehat {w}  = 0
		 		\label{cont_FT1}
		 	\end{equation}
		 	Equations (\ref{mom_perp_FT}), (\ref{mom_z_FT}) and (\ref{cont_FT1}) represent the FFT of the incompressible NS equations expressed in EC notation. Next, we will proceed to discretize these equations using DEC.

		 	
		 First, the continuity equation is discretized as
		 	\begin{equation}
		 		\ast_2 \  d_1 \ \ast^{-1}_1 \ \widehat {U} 	+  i \ k \ \widehat {W}  = 0
	\label{cont_FTdisc}	 	
    \end{equation}
		 	where $U$ and  $W$ represent the discrete dual velocity 1-form and discrete dual velocity 0-form respectively. Let $U = \ast_1 \ U^{\ast}$ and substitute the identity $\ast^{-1}_1  \ast_1 = -1$ in Equation (\ref{cont_FTdisc}) yields discretized continuity equation as
		 	\begin{equation}
		 		- \ast_2 \  d_1 \ \widehat {U^\ast} 	+  i \ k \ \widehat {W}  = 0
		 		\label{cont}
		 	\end{equation}
		 	
Discretizing $\perp$- momentum Equation (\ref{mom_perp_FT}) 
	\begin{equation}
	\deriv[]{\widehat{\uperp}}{t} + \ast \reallywidehat{ (\uperp \wedge \ast d \uperp) } + i k \reallywidehat{ w \wedge \uperp } = - d \widehat{p} -  \frac{1}{2} d \widehat{|\uperpv|^2} + \mu \Big(  \ast d \ast d \widehat{\uperp} - i k d \widehat{w} - k^2 \widehat {u^{\perp}}      \Big)
	\label{mom_perp_FT}
	\end{equation}
\begin{center}
 \Large $\Downarrow$
  \end{center}
	\begin{equation}
	\begin{aligned}
	\frac{\widehat{U^{n+1}} - \widehat{U^{n}}}{\Delta t} 
	+ \ast_1 \reallywidehat{  M_v \ast^{-1}_0 ( [-d^{T}_0] U + d_b V  )  }
	+ i k \widehat{ M_u W }	+ d^{T}_1 \widehat{P} + \frac{1}{2} \ d^{T}_1 \ \widehat{|\uperpv|^2} \\
	-\mu \Big(   \ast_1 d_0 \ast^{-1}_0 ( [-d^{T}_0] \widehat{U} + d_b \widehat{V}  )  - i k d^{T}_1 \widehat{W} - k^2 \widehat{U}           \Big) 
	= 0
	\end{aligned}
	\label{mom2}
	\end{equation}

where $U$, $V$, $W$ and $P$ are vectors
\begin{itemize}
\item $V$: the 1-forms representing tangential velocities on all edges of the primal mesh 
\item $P$:  the 0-forms representing pressure at all vertices of the dual mesh
\item $[-d^{T}_0]U$: calculates the circulation of velocity forms $u$ along the boundaries of dual $2$-cells
\item $d_bV$: augments the velocity circulation by addressing cases where dual 2-cell boundaries include primal edges, accounting for contributions from velocity 1-forms $v$ on these primal boundary edges
 \item $M_v = \frac{1}{2} V |d_0| $ : product formed by the wedge operation involving a primal 1-form $v^{\perp}$ and 0-form $\ast d u^{\perp}$
 \item $M_u = \frac{1}{2} U |d^{T}_1| $ : wedge product between dual 1-form $u^{\perp}$ and dual 0-form $w$
\item $\Delta t$ : discrete time step between the current time $n + 1$ and the previous time $n$.
\end{itemize}

	Let $\widehat{U} = \ast_1 \widehat{U^{\ast}}$, Equation (\ref{mom2}) becomes
		
		\begin{equation}
		\begin{aligned}
	&	\ast_1 \ \frac{\widehat{{U^\ast}^{n+1}} - \widehat{{U^\ast}^{n}}}{\Delta t} 
		+ \ast_1 \ \reallywidehat{ M_v \ast^{-1}_0 ( [-d^{T}_0] \ast_1 U^{\ast} + d_b V  )  }\\
	&	+ i k \widehat{ M_u W }
		+ d^{T}_1 \widehat{P}  
        + \frac{1}{2} d^{T}_1 \widehat{|\uperpv|^2}\\
	&	-\mu \Big(   \ast_1 d_0 \ast^{-1}_0 ( [-d^{T}_0] \ast_1 \widehat{U^{\ast}} + d_b \widehat{V}  )  - i k d^{T}_1 \widehat{W} - \ast_1 k^2 \widehat{U^{\ast}} \Big) 
		= 0
		\end{aligned}
		\label{mom3}
		\end{equation}

  	Applying $\ast^{-1}_1$ and $\ast^{-1}_1 \ \ast_1 = -1$
	
	\begin{equation}
	\begin{aligned}
	&	- \frac{\widehat{{U^\ast}^{n+1}} - \widehat{{U^\ast}^{n}}}{\Delta t} 
		-  \reallywidehat{ M_v \ast^{-1}_0 ( [-d^{T}_0] \ast_1 U^{\ast} + d_b V  )  }\\
	&	+ i k \ast^{-1}_1 \widehat{ M_u W }
		+ \ast^{-1}_1 d^{T}_1 \widehat{P} 
  + \frac{1}{2} \ast^{-1}_1 d^{T}_1 \widehat{|\uperpv|^2}\\
	&	-\mu \Big(  - d_0 \ast^{-1}_0 ( [-d^{T}_0] \ast_1 \widehat{U^{\ast}} + d_b \widehat{V}  )  - i k \ast^{-1}_1 d^{T}_1 \widehat{W} + k^2 \widehat{U^{\ast}} \Big) 
		= 0
		\end{aligned}
	\label{pmom}
	\end{equation}
		 		 	
		 	Discretizing the z-momentum Equation (\ref{mom_z_FT})
	\begin{equation}
	\frac{\widehat{W^{n+1}} - \widehat{W^{n}}}{\Delta t} 
	+ \ast_2 d_1 \ast^{-1}_1 \widehat{M_u W} + i k \widehat{W^2} 
	+ i k \widehat {P}
	- \mu \Big(  \ast_2 d_1 \ast^{-1}_1 d^{T}_1 \widehat{W} - k^2 \widehat{W}    \Big) 
	= 0
	\label{zmom}
	\end{equation}
This concludes the discretization process and a linear system can be formulated by combining Equations (\ref{pmom}), (\ref{zmom}) and (\ref{cont}) together as follows
	\begin{equation}
	\begin{aligned}
	\bigg[- \frac{1}{\Delta t}I + \mu d_0\ast^{-1}_0 [-d^{T}_0] \ast_1	\bigg] \widehat{{(U^\ast)}^{n+1}} +\frac{1}{2} \ast^{-1}_1 d^{T}_1 \widehat{(P)^{n+1}}  - \mu ik\ast^{-1}_1 d^{T}_1 \widehat{(W)^{n+1}}= F_1 
	\end{aligned}
	\label{momf}
	\end{equation}
	\begin{math}
	\begin{aligned}
	F_1 = \big[ \mu k^2 - \frac{1}{\Delta t}\big] \widehat{({U^\ast})^{n}} - \mu  d_0 \ast^{-1}_0 d_b  \widehat{(V)^n} +  \frac{1}{2} \ast^{-1}_1 d^{T}_1 \ast_2 \reallywidehat{(M_{v1} U^{\ast})^n  } \\
 + i k \ast^{-1}_1  \reallywidehat{(M_u W)^n} -\frac{1}{2} \ast^{-1}_1 d^{T}_1 \widehat{(P)^{n}}
	\end{aligned}
	\label{momf1}
	\end{math}
	\begin{equation}
	\big[0 \big] \widehat{{(U^\ast)}^{n+1}} + \frac{1}{2} ikI\widehat{(P)^{n+1}} + \bigg[ \frac{1}{\Delta t}I - \mu \ast_2 d_1 \ast^{-1}_1 d^{T}_1 \bigg] \widehat{(W)^{n+1}} = F_2
	\label{zmomf}
	\end{equation}
	\begin{math}
	F_2 = \big[ \frac{1}{\Delta t}I - \mu k^2 I \big] \widehat{(W)^{n}} - ik \ I \widehat{(W^2)^{n}} - \ast_2 d_1 \ast^{-1}_1 \widehat{(M_u W)^n} -\frac{1}{2} ikI\widehat{(P)^{n}}
	\label{zmomf2}
	\end{math}	
	
	\begin{equation}
	\big[- \ast_2  d_1\big] \widehat{{(U^\ast)}^{n+1}} +  \big[0\big] \widehat {(P)^{n+1}} 	+  \big[ik \ I\big] \widehat {(W)^{n+1}}  = 0
	\label{contf}
	\end{equation}
	
Equations (\ref{momf}), (\ref{zmomf}), and (\ref{contf}) constitute a linear system with $\widehat{U^{\ast}}$, $\reallywidehat{P}$, and $\widehat{W}$ as the degrees of freedom, where $I$ denotes the identity matrix.




\section{Problem Formulation \label{LDCProblemFormulation}}
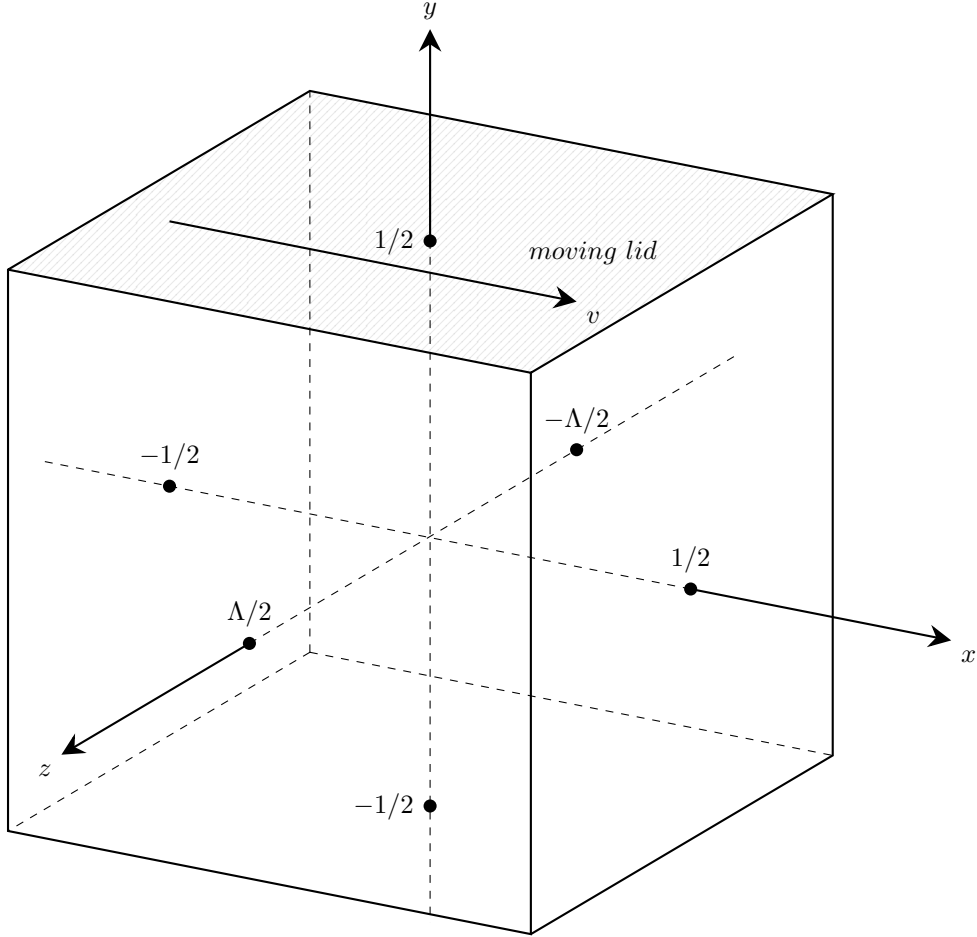
\begin{figure}[h]
	\centering
\begin{tikzpicture}[tdplot_main_coords]
\def\BigSide{8}
\def\SmallSide{2}
\pgfmathsetmacro{\CalcSide}{\BigSide-\SmallSide}

\tdplotsetcoord{P}{sqrt(3)*\BigSide}{55}{45}

\coordinate (sxl) at (\BigSide,\CalcSide,\BigSide);
\coordinate (syl) at (\CalcSide,\CalcSide,\BigSide);
\coordinate (szl) at (\CalcSide,\BigSide,\BigSide);
\coordinate (zp) at (12,6,5);
\coordinate (zm) at (-13,0,0);
\coordinate (xp) at (2,7,2.8);
\coordinate (xm) at (0,-10,0);
\coordinate (yp) at (2,3,7);
\coordinate (ym) at (0,0,-9.65);
\coordinate (xml) at (2,-1,2.8);
\coordinate (yml) at (2,3,-1);
\coordinate (zml) at (3.3,6,5);

\draw[dashed] 
  (0,0,0) -- (Px)
  (0,0,0) -- (Py)
  (0,0,0) -- (Pz);
\draw[-{Stealth[length=3mm, width=3mm]}] [thick]
  (zp) -- ++ (5,0,0) node[anchor=north east]{$z$};
\draw[-{Stealth[length=3mm, width=3mm]}] [thick]
   (xp) -- ++(0,4,0) node[anchor=north west]{$x$};
\draw[-{Stealth[length=3mm, width=3mm]}] [thick]
  (yp) -- ++(0,0,3) node[anchor=south]{$y$};

\draw[dashed]
  (zp) -- ++ (zm) ;
\draw[dashed]
  (xp) -- ++ (xm) ;
\draw[dashed]
  (yp) -- ++ (ym) ;
  
\draw[thick]
  (Pxz) -- (P) -- (Pxy) -- (Px) -- (Pxz) -- (Pz) -- (Pyz) -- (P); 
\draw[thick]
  (Pyz) -- (Py) -- (Pxy);

\node[label=above:$1/2$,fill,circle,inner sep=1.75pt] at (xp) {};
\node[label=left:$1/2$,fill,circle,inner sep=1.75pt] at (yp) {};
\node[label=above:$\Lambda/2$,fill,circle,inner sep=1.75pt] at (zp) {};

\node[label=above:$-1/2$,fill,circle,inner sep=1.75pt] at (xml) {};
\node[label=left:$-1/2$,fill,circle,inner sep=1.75pt] at (yml) {};
\node[label=above:$-\Lambda/2$,fill,circle,inner sep=1.75pt] at (zml) {};

\fill[pattern=north east lines,opacity=0.3]
    (Pxz)  -- (Pz) -- (Pyz) -- (P); 

\draw[-{Stealth[length=3mm, width=3mm]}] [thick]
  (2,-1,6.55) -- ++ (0,6.25,0) node[anchor=north west]{$v$};

\draw[]
  (2,5.5,7)  node[anchor=south]{$moving$ $lid$};

\end{tikzpicture}
    \caption{Geometry of the $three{\text -}dimensional$ lid-driven cavity.}
    \label{topology3D}
\end{figure}

The lid-driven cavity (LDC) problem holds significant importance as a benchmark test in validating numerical methods for fluid flow simulations. This classic problem serves as a rigorous and widely recognized standard, facilitating the assessment of the accuracy and reliability of numerical simulations of fluid dynamics. Various numerical methods have been widely employed to comprehensively establish the solution for the lid-driven cavity problem through the primitive variable formulation \cite{schreiber1983driven} or by the vorticity and stream function equations \cite{ghia1982high}. The solution to the problem is composed of a primary large eddy, which is the dominant in the flow structure, and three secondary eddies positioned in close proximity to the corners of the cavity. This arrangement of eddies characterizes the complex flow patterns within the system, highlighting the interplay and dynamics between the primary and secondary features. The primary large eddy signifies the main rotational motion, while the secondary eddies contribute to additional vortical structures, collectively shaping the overall fluid behavior in the lid-driven cavity problem. The system also demonstrates a specific singularity at the boundary conditions where moving and stationary walls intersect \cite{gupta1981nature,hancock1981effects}. The configuration of the 3D lid-driven cavity is illustrated in Figure \ref{topology3D}. The $xy$ plane represents the flat, 2D surface formed by the intersection of the $x{\text -}$axis and $y{\text -}$axis, while the span-wise dimension extends along the $z{\text -}$axis, signifying the third dimension in a 3D space. The geometry is characterized by the span-wise aspect ratio $\Lambda = \frac{l_z}{L}$ where $L$ defines the length of the cavity in both $x{\text -}$ and $y{\text -}$directions and $l_z$ is the domain length in the $z{\text -}$direction. The boundary conditions in the $x{\text -}$ and $y{\text -}$directions are no-slip boundary condition as follows:\\
for the normal velocity
\begin{subequations}
  \begin{align}
& u(x= \pm 1/2)= 0 \\
& u(y= \pm 1/2)= 0 
  \end{align}
\end{subequations}
\\
and the tangential velocity
\begin{subequations}
  \begin{align}
& v(y=  1/2)= 1 \\
& v(y= - 1/2)= 0 
  \end{align}
\end{subequations}
meanwhile, the periodic boundary conditions are imposed at $z = \pm \Lambda/2$:
\begin{subequations}
  \begin{align}
& u(z=\Lambda/2) = u(z=-\Lambda/2) \\
& v(z=\Lambda/2) = v(z=-\Lambda/2) \\
\end{align}
\end{subequations}

\tdplotsetmaincoords{70}{120}

	\begin{figure}[h]
		\centering
		\includegraphics[height=4.5in]{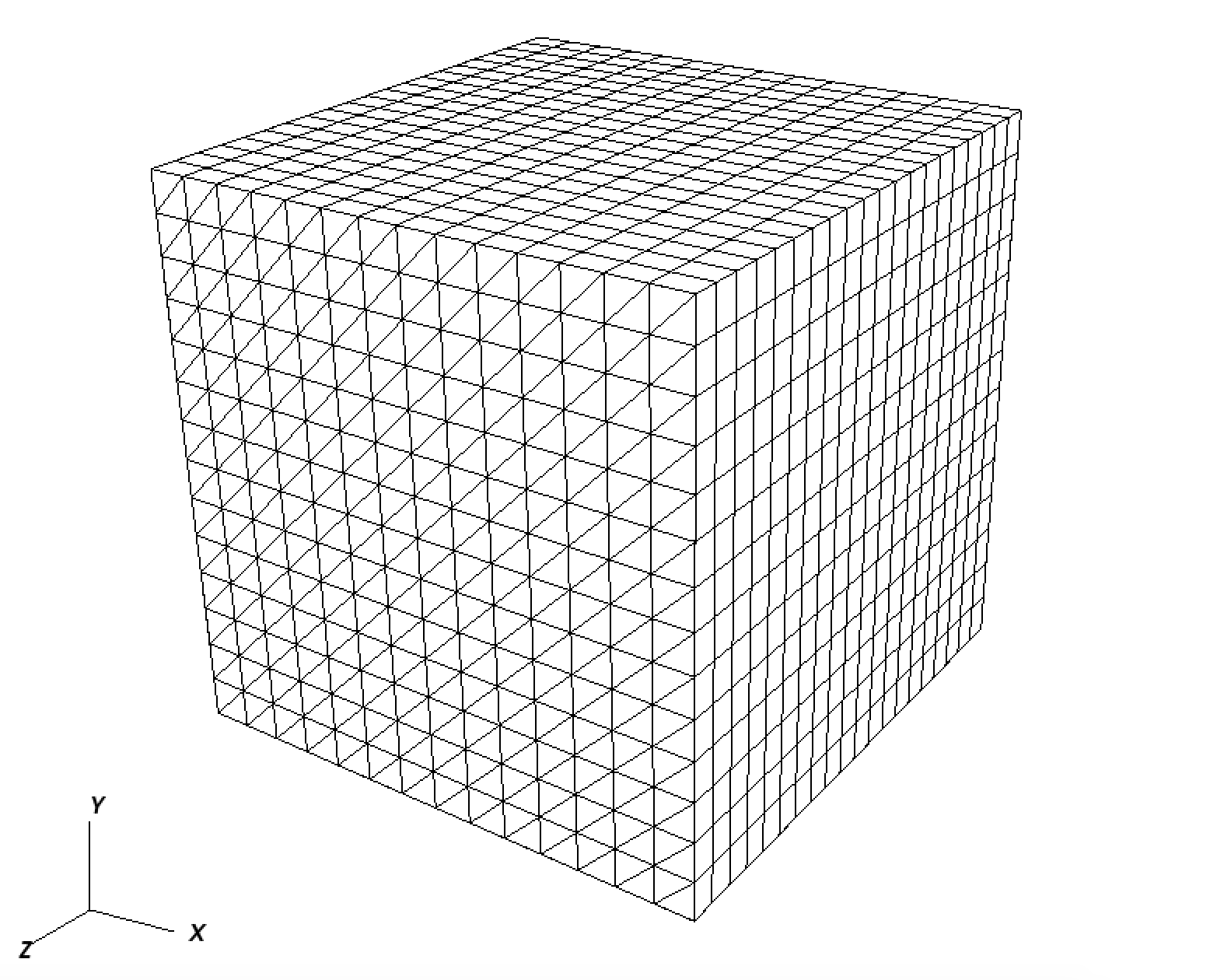}
          \caption{Topology of 3D mesh resolution of $28 \times 28 \times 20$}
          \label{mesh3D}
	\end{figure}

\section{Validation of the Code \label{LDCValidation}}
\begin{figure}[h]
		\centering
		\includegraphics{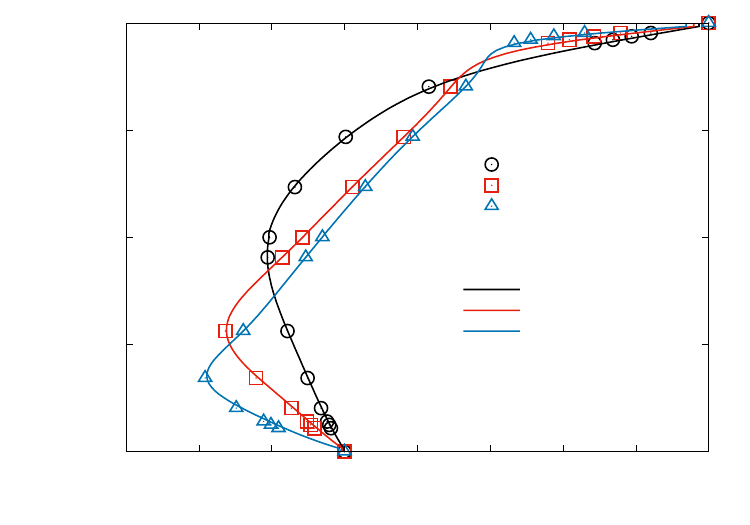}
		\caption{Profile of the $x{\text -}$component velocity, $u_x$ along the vertical centerline of a square unit cavity for different Reynolds numbers compared to the findings of Ghia et al. \cite{ghia1982high}. The resolution is $N_x \times N_y = 256 \times 256$.}
		\label{2dghia_u}
	\end{figure}

\begin{figure}[h]
		\centering
		\includegraphics{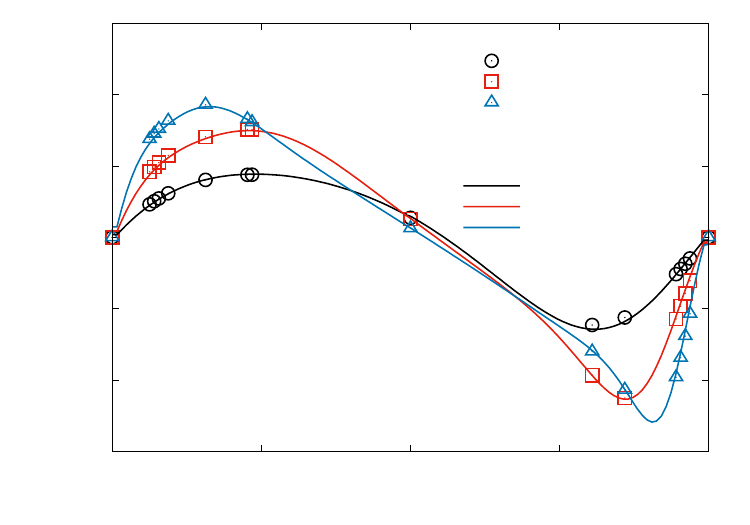}
		\caption{Profile of the $y{\text -}$component velocity, $u_y$ along the horizontal centerline of a square unit cavity for different Reynolds numbers compared to the findings of Ghia et al. \cite{ghia1982high}. The resolution is $N_x \times N_y = 256 \times 256$.}
		\label{2dghia_v}
	\end{figure}

The validation process serves as a crucial step in ensuring the accuracy and reliability of the implemented code by benchmarking it against established and validated results from literature. As an initial step, it is imperative to conduct code validation utilizing the $two{\text -}dimensional$ (2D) square lid-driven cavity flow. In this test case, the span-wise velocity $w$ is specifically set to zero. The simulations of a driven cavity are conducted on a unit square domain for Reynolds numbers $(Re)$ of $100$, $400$ and $1000$. Solid wall boundary conditions are applied to the left, right, and bottom boundaries. The top boundary features zero flux for $u$ and unit tangential velocity for $v$, with a time step of $\Delta t = 0.01$. These simulations are executed on a structured-triangular mesh, comprising isosceles right triangles in the $xy$ plane as shown in Figure \ref{mesh3D}. While in the $z{\text -}$direction, prismatic elements are considered to discretize the domain, aligning with the use of FFT discretization in the span-wise direction.
The obtained results from the simulations are meticulously compared against the findings presented by Ghia et al. \cite{ghia1982high}, for the unit square cavity for the various aforementioned $Re$ values. In Figures \ref{2dghia_u} and \ref{2dghia_v}, the graphs depict profiles of both horizontal velocity, $u_x$ along the geometrical vertical centerline and vertical velocity, $u_y$ along the geometrical horizontal centerline for all mentioned Reynolds numbers at simulation time $T = 100$. The outcomes from our study align closely with those published by Ghia et al. \cite{ghia1982high}, demonstrating consistent agreement. This leads to confidence in the accuracy and reliability of the present approach.

\begin{figure}[h]
  \centering
  \includegraphics[height=9in]{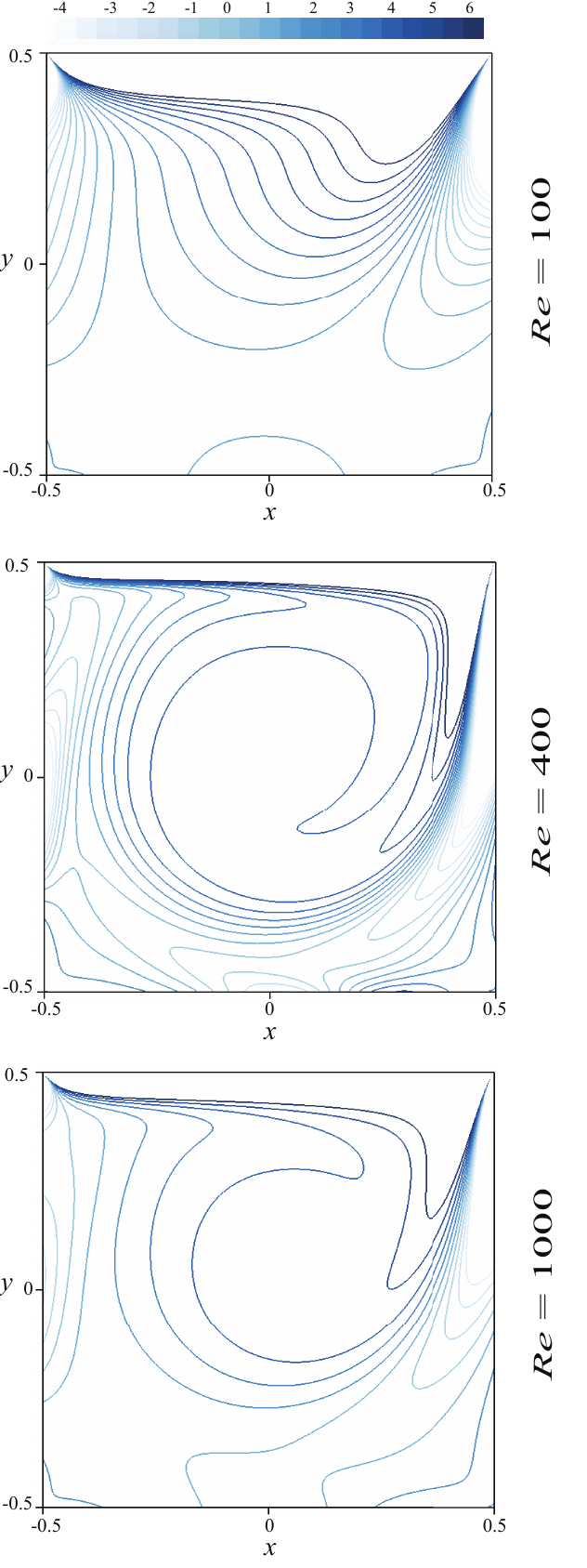}
  \caption{Vorticity contours of present study for 2D cavity flow for different Reynolds numbers}.
  \label{vorticitycontours}
\end{figure}

\begin{figure}[h]
		\centering
		\includegraphics{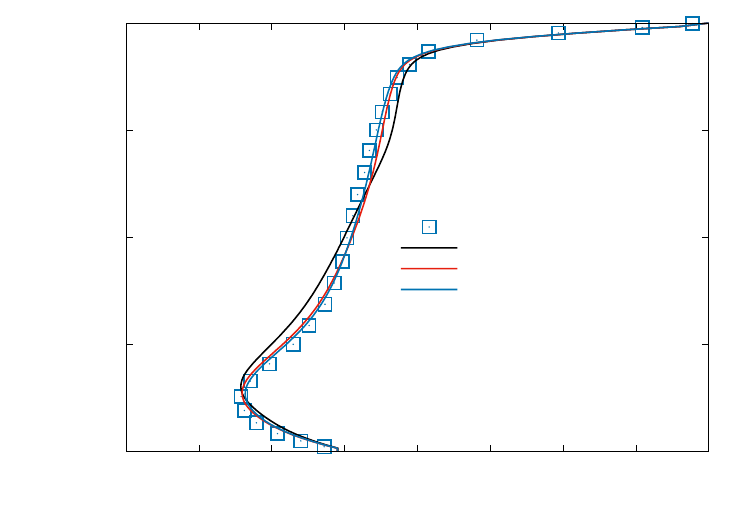}
		\caption{Profile of the $x{\text -}$component velocity, $u_x(0,y,0)$ along the vertical center line of a cubic unit cavity (i.e, $\Lambda = 1$) for $Re = 1000$ for varying spatial resolution in the span-wise direction $N_z$ compared to the findings of Ku et al.\cite{ku1987pseudospectral}}		
		\label{3dku_u}
	\end{figure}

\begin{figure}[h]
		\centering
		\includegraphics[height=4.5in]{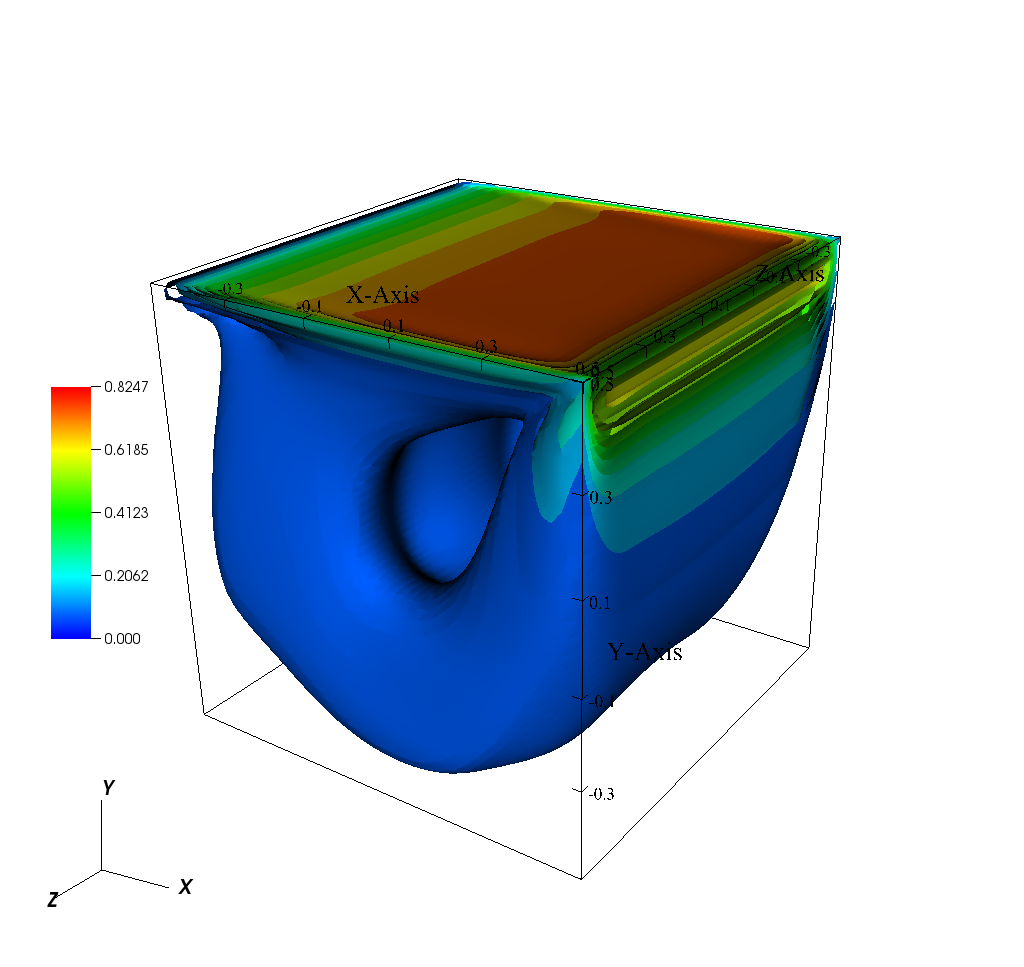}
          \caption{Isosurface of the velocity magnitude of a cubical cavity for $Re=1000$ and $\Lambda=1$ obtained by mesh resolution of $96 \times 96 \times 64$}.	\label{vmagisosurf96}
	\end{figure}

\begin{figure}[h]
		\centering
		\includegraphics[height=4.5in]{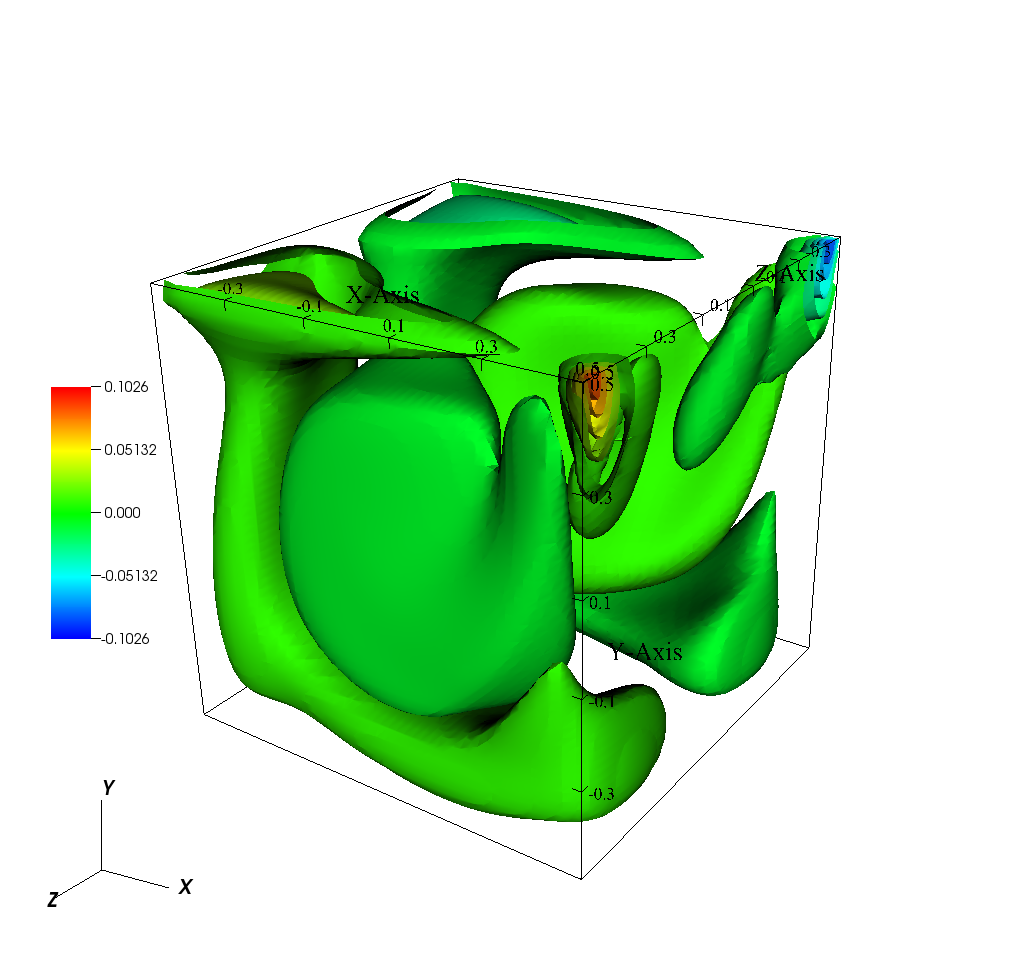}
          \caption{Isosurface of the $z{\text -}$component velocity, $u_z$ of a cubical cavity for $Re=1000$ and $\Lambda=1$ obtained by mesh resolution of $96 \times 96 \times 64$}.
          \label{omegaz96x64}
	\end{figure}

The contour plots of the z-component vorticity, $\omega_z$ at different Reynolds numbers are depicted in Figure \ref{vorticitycontours}. As the Reynolds number Re increases, the vorticity contours predominantly highlight the primary vortex (PV). However, they do not disclose the existence of smaller secondary vortices that emerge at the corners of the cavity. Notably, the results exhibit a commendable agreement with those published by Ghia et al. \cite{ghia1982high}.
 \\
In order to validate the $three{\text -}dimensional$ hybrid DEC-FFT method, the simulations are carried out for 3D lid-driven cavity problem at different Reynolds numbers Re using three different mesh resolutions $256 \times 256 \times 64$, $256 \times 256 \times 16$ and $256 \times 256 \times 32$. Notably, the spatial resolution $N_x$ and $N_y$ are fixed, while the spatial resolution in the $z{\text -}$direction, $N_z$ is varying to evaluate spectral accuracy using FFT in the third dimension. Figures \ref{3dku_u} and \ref{3dku_v} illustrate the profiles of the $x{\text -}$component velocity, $u_x(0,y,0)$ along the vertical center line and the $y{\text -}$component velocity, $u_y(x,0,0)$ along the horizontal center line of a cubic unit cavity (i.e, $\Lambda = 1$). Simulations are carried out for $Re = 1000$ for varying spatial resolution in the span-wise direction $N_z$ compared to the findings of Ku et al.\cite{ku1987pseudospectral}. By increasing the spectral resolution in the third dimension, the method yields velocity profiles that closely align with those found in the literature by Ku et al. \cite{ku1987pseudospectral}. The Iso-surfaces of the velocity magnitude and the $z{\text -}$component velocity, $u_z$ are shown in Figures \ref{vmagisosurf96} and \ref{omegaz96x64}, respectively.

	\begin{figure}[h]
		\centering
		\includegraphics{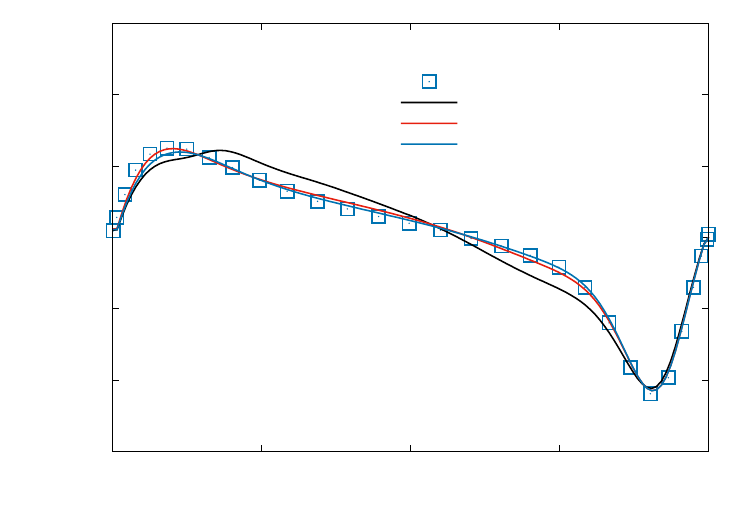}
		\caption{Profile of the $y{\text -}$component velocity, $u_y(x,0,0)$ along the horizontal center line of a cubic unit cavity (i.e, $\Lambda = 1$) for $Re = 1000$ for varying spatial resolution in the span-wise direction $N_z$ compared to the findings of Ku et al.\cite{ku1987pseudospectral}}		
  \label{3dku_v}
	\end{figure}
	\begin{figure}[h]
		\centering
		\includegraphics{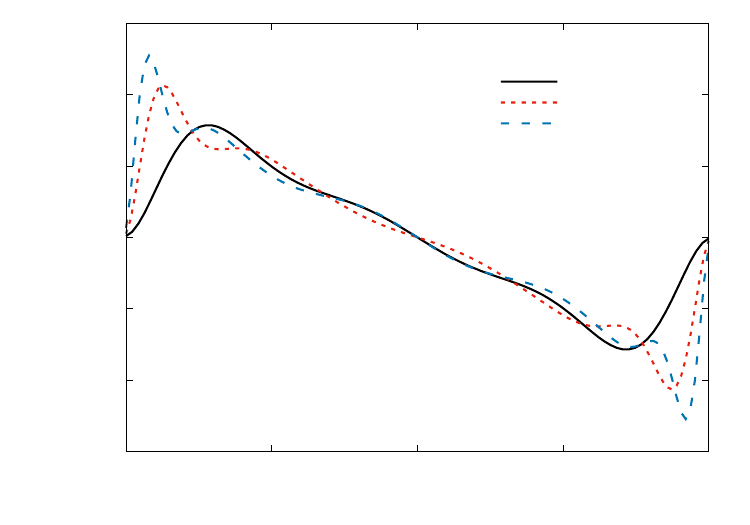}
		\caption{Profile of the $z{\text -}$component velocity, $w(0,0,z)$ along the spanwise direction for different aspect ratios (i.e, $\Lambda = 1,2,$ and $3$) for $Re = 1000$. The resolution is $128 \times 128 \times 96$ for $\Lambda = 1$ and $128 \times 128 \times 128$ for $\Lambda = 2,3$.}		
  \label{3dku_v}
	\end{figure}

The Taylor-Green Vortex (TGV) is widely recognized as a significant benchmark problem in computational fluid dynamics. Its importance stems from its role as a standard test case utilized to validate and assess numerical solvers and simulation methods. It was first introduced by Taylor and Green \cite{taylor1937mechanism} to interpret the formation of small scale eddies from large ones through vortex-stretching, diffusion, and dissipation in a $three{\text -}dimensional$ (3D) flow domain.\\

\section{Two Dimensional Taylor-Green Vortices \label{2DTGV}}
In the first test, the Taylor–Green vortices are simulated in 2D in a square domain with dimensions $[-\pi,\pi]$ in both $x{\text -}$, and $y{\text -}$directions. The initial velocity components at $T=0$ in $x{\text -}$, and $y{\text -}$directions are given by:\\
\begin{subequations}
  \begin{align}
& u_x(x,y,0) = cos(x) sin(y) \\
& u_y(x,y,0) = -sin(x) cos(y)
  \end{align}
\end{subequations}

\begin{figure}
		\centering
		\includegraphics[width=0.75\columnwidth]{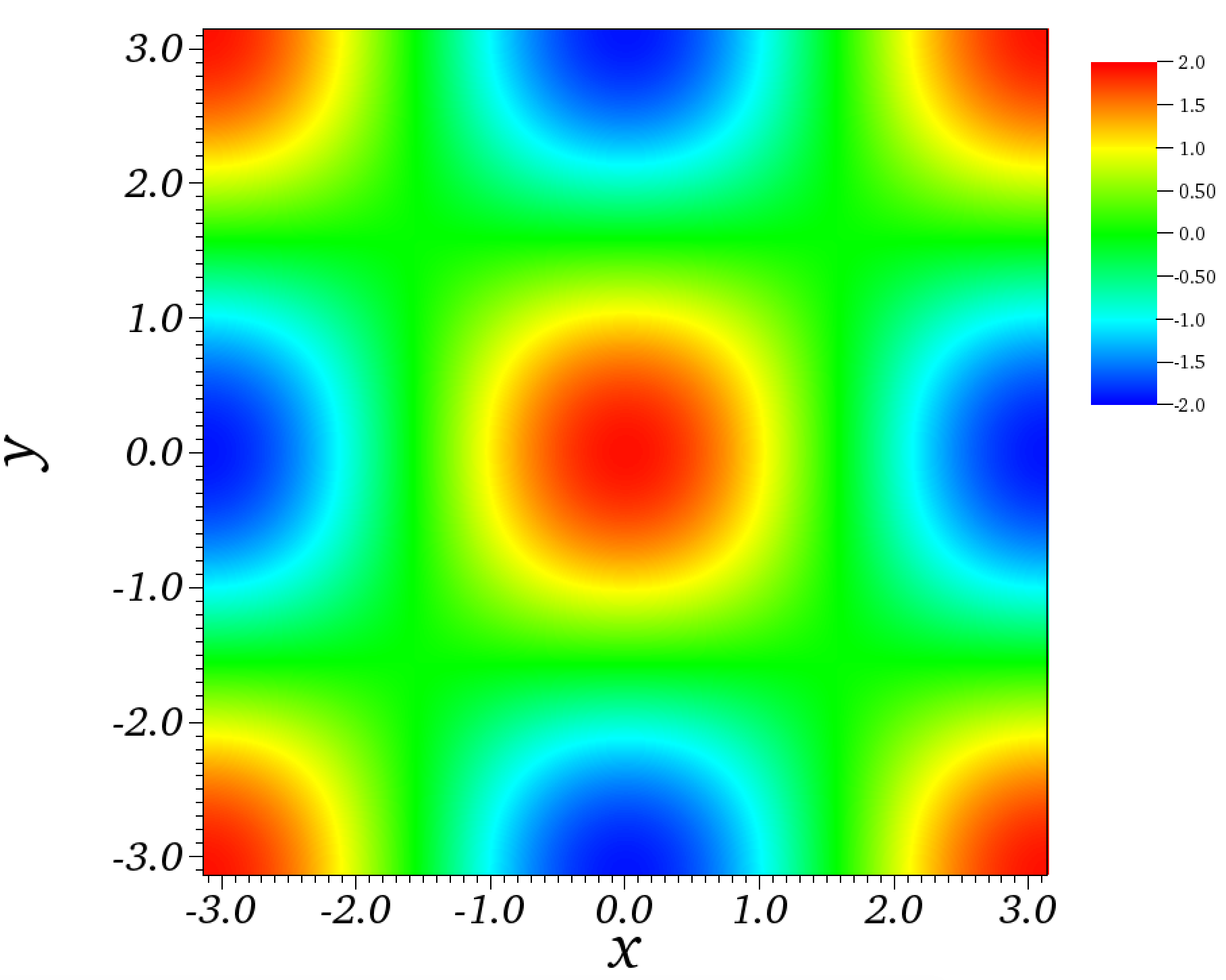}
		\caption{A contour plot depicting vorticity for Taylor-Green vortices at time $T=10$.}
		\label{TGV-vorticity2D}
	\end{figure}

\begin{figure}[!]
		\centering
		\includegraphics{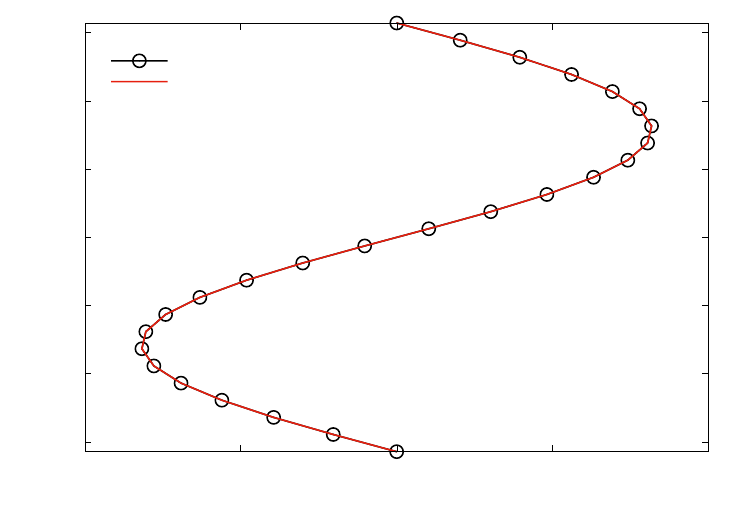}
		\caption{The profile of the $x$-component velocity, $u_x$, along the vertical center line for Taylor-Green vortices at time $T=10$.}
		\label{2dTGV_ux}
	\end{figure}

\begin{figure}[!]
		\centering
		\includegraphics{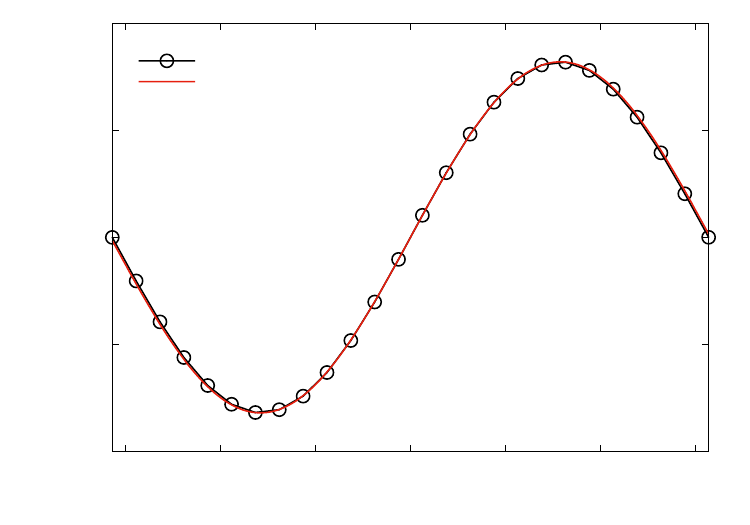}
		\caption{The profile of the $y$-component velocity, $u_y$, along the horizontal center line for Taylor-Green vortices at time $T=10$.}
		\label{2dTGV_uy}
	\end{figure}

The decay of Taylor–Green vortices over time possesses an analytical solution, which is explicitly expressed in the case of 2D dynamics in \cite{taylor1937mechanism,connors2010convergence} as\\
\begin{subequations}
  \begin{align}
& u_x = -cos(x) sin(y) e^{-2\nu t}\\
& u_y = sin(x) cos(y) e^{-2\nu t}
  \end{align}
\end{subequations}
where $\nu$ denotes the kinematic viscosity. A mesh resolution of $256 \times 256$ is considered for the simulation with periodic boundary conditions applied to all boundaries of the domain. The simulation incorporates a time step of $\Delta t = 0.01$, with a kinematic viscosity of $\nu = 0.01$. 
Figure \ref{TGV-vorticity2D} illustrates the vorticity contour plot of the Taylor-Green vortices at a simulation time of $T=10$. The computed velocity profile of both $x{\text -}$component, $u_x$ and $y{\text -}$component, $u_y$ along the domain vertical center line and horzintal line are plotted along with those of analytical values in Figure \ref{2dTGV_ux} and Figure \ref{2dTGV_uy}, respectively. The computed velocities exhibit strong conformity with the analytical solution. This offers a qualitative indication of the reliability of the current numerical implementation in replicating the time evolution of such unsteady flow.


\section{Three Dimensional Taylor-Green Vortices \label{3DTGV}}
\begin{figure}[!]
		\centering
		\includegraphics[width=1.0\columnwidth]{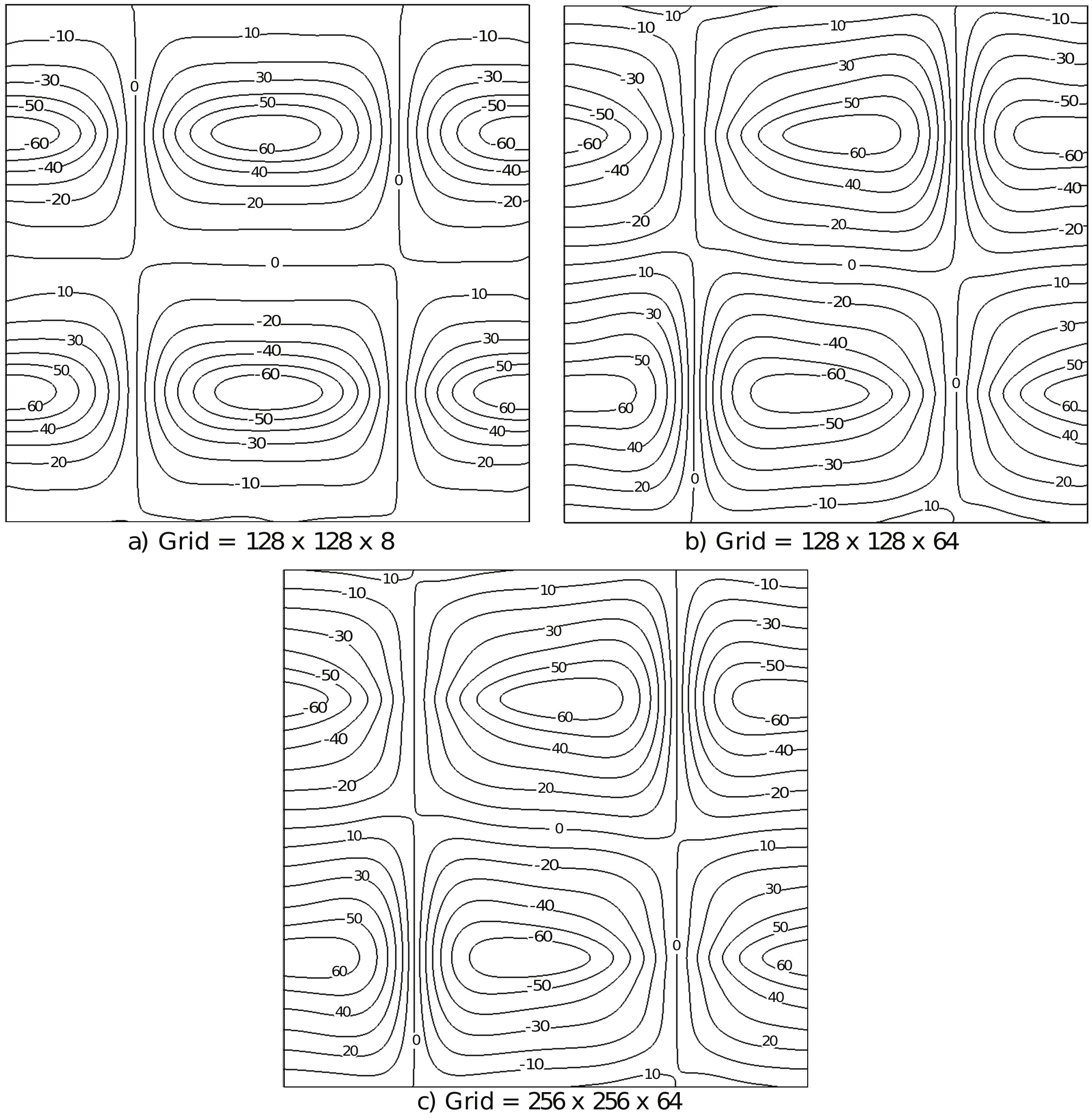}
 \caption{Contour plots of $x{\text -}$component velocity, $u_x$ obtained by different spatial resolution at $z = \pi/4{\text -}$plane, $Re = 100$ at simulation time $T = 3.5$.}
 \label{ux3D_TGV}
 \end{figure}

\begin{figure}[!]
		\centering
		\includegraphics[width=1.0\columnwidth]{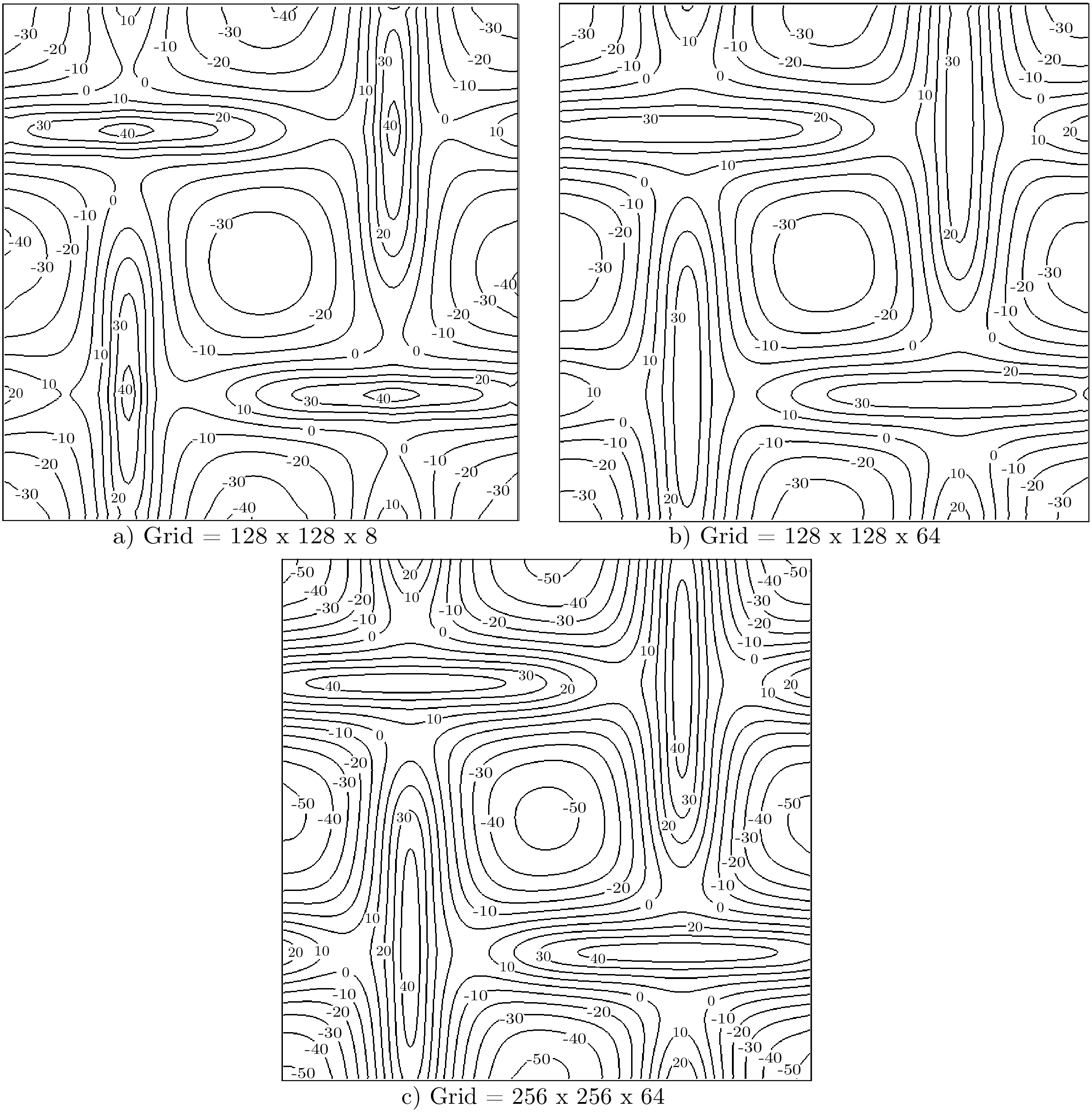}
 \caption{Contour plots of $z{\text -}$component velocity, $u_z$ obtained by different spatial resolution at $z = \pi/4{\text -}$plane, $Re = 100$ at simulation time $T = 3.5$.}
 \label{uz3D_TGV}
 \end{figure}

In this test, the hybrid DEC-FFT is tested for the Taylor–Green vortices in 3D domain. The simulations are carried out in a cube with dimensions $[-\pi,\pi]$ in all $x{\text -}$, $y{\text -}$ and $z{\text -}$directions. The initial velocity components at $T=0$ in $x{\text -}$, $y{\text -}$ and $z{\text -}$directions are given by:\\
\begin{subequations}
  \begin{align}
& u_x(x,y,z,0) = cos(x) sin(y) cos(z) \\
& u_y(x,y,z,0) = -sin(x) cos(y) cos(z) \\
& u_z(x,y,z,0) = 0
  \end{align}
  \label{initial3dTGV}
\end{subequations}
In Figure \ref{ux3D_TGV}, contour plots of the $x{\text -}$component velocity, $u_x$ at simulation time $T = 3.5$, $Re = 100$ at $z = \pi/4{\text -}$plane obtained by different meshes of varying spatial resolutions. These plots depict the nature of the flow that evolves from the initial conditions (\ref{initial3dTGV}).\\
Despite the initial condition of having zero $z{\text -}$component velocity, the flow that evolves from \ref{initial3dTGV} is $three{\text -}dimensional$. This is illustrated in Figure \ref{uz3D_TGV} by the contour plots of the $z{\text -}$component velocity, $u_z$ at simulation time $T = 3.5$, $Re = 100$ at $z = \pi/4{\text -}$plane obtained by different meshes of varying spatial resolutions. These qualitative findings align well with those presented by Orszag \cite{orszag1974numerical} and Sharma and Sengupta \cite{sharma2019vorticity}.


\section{Summary and Directions for the Future Work \label{Summary}}
The paper discusses the importance of simulating fluid flow problems, particularly incompressible flows governed by the Navier-Stokes equations, across diverse scientific and engineering applications. Conventional numerical methods for solving these equations on three-dimensional meshes are recognized for their moderate conservation characteristics and significant computational demand. To address these issues, the paper proposes a hybrid discretization method for solving the 3D incompressible Navier-Stokes equations using discrete exterior calculus and Fourier transform. The proposed hybrid DEC-FFT method aims to combine the advantages of both discrete exterior calculus and Fourier transform to achieve accurate and efficient simulations of fluid flow problems. The paper presents a hybrid discretization approach for solving the 3D incompressible Navier-Stokes equations using discrete exterior calculus and Fourier transform. The paper discusses the theoretical foundations of both discrete exterior calculus and Fourier transform and how they can be integrated to solve fluid flow problems. The paper also introduces an existing conservative primitive variable DEC discretization method developed by Jagad et al. \cite{jagad2021primitive}. The paper then extends this method by incorporating the Fourier transform technique to handle frequency components and improve computational efficiency. The hybrid DEC-FFT method is validated and tested on two benchmark problems, namely lid-driven cavity flow and simulation of Taylor-Green vortices. The results of the simulations using the hybrid DEC-FFT method are compared with literature results, demonstrating the accuracy and effectiveness of the proposed approach. In conclusion, the paper presents a hybrid discretization approach that combines discrete exterior calculus and Fourier transform for solving the 3D incompressible Navier-Stokes equations. The proposed hybrid DEC-FFT method offers improved conservation properties and computational efficiency compared to traditional numerical methods. Furthermore, the method is able to accurately capture the geometric and topological properties of the mesh, making it suitable for complex flow simulations which leads to accurate and efficient simulations of fluid flow problems. 



\section*{Acknowledgments}
This research was supported by KACST and the KAUST Office of Sponsored Research under Award BAS/1/1424-01-01.
\begin{onehalfspacing}
	\renewcommand*\bibname{\centerline{REFERENCES}} 
    \phantomsection
	\addcontentsline{toc}{chapter}{References}
	\newcommand{\BIBdecl}{\setlength{\itemsep}{0pt}}
    \printbibliography
    
    %
    %
    \begingroup
    \makeatletter
    \@ifundefined{ver@biblatex.sty}
    {\@latex@error
    	{Missing 'biblatex' package}
    	{The bibliography requires the 'biblatex' package.}
    	\aftergroup\endinput}
    {}
    \endgroup

\end{onehalfspacing} 






\end{document}